\documentclass[twocolumn]{aastex701}

\usepackage{xcolor}
\usepackage{amsmath,amssymb,bm}
\usepackage{mathtools}
\usepackage{booktabs}
\usepackage{enumitem}
\usepackage{hyperref}
\usepackage{svg}

\usepackage{booktabs}
\usepackage{tabularx}
\usepackage{array}

\usepackage{graphicx}
\usepackage{subcaption}

\newcommand{\sech}{\operatorname{sech}}
\newcommand{\RLC}{R_{\rm LC}}

\newcommand{\Omi}{\Omega_i}

\newcommand{\Deltap}{\Delta}
\newcommand{\Ciprof}{\mathcal{C}_i}
\newcommand{\Cpprof}{\mathcal{C}_p}

\begin{document}

\shorttitle{Ion-acoustic-like modes in ion-loaded pulsar current sheets}
\shortauthors{Singh et al.}

\title{Ion-Acoustic-Like Modes in Ion-Loaded Pulsar-Wind Current Sheets: A Pressure-Balanced Existence Criterion}

\author[0000-0003-0289-2818]{Manpreet Singh}

\affiliation{School of Computing and Artificial Intelligence, Southwest Jiaotong University, Chengdu-610031, PR China.}
\affiliation{School of Physical Science and Technology, Southwest Jiaotong University, Chengdu-610031, PR China}
\email[show]{singhmanpreet185@gmail.com}

\author[0000-0003-3855-4968]{Ripin Kohli}
\affiliation{Department of Applied Sciences, Khalsa College of Engineering and Technology, Amritsar-143001, Punjab, India}
\email[]{ripin.kohli@gmail.com}

\author[0000-0003-1039-9521]{Siming Liu}
\affiliation{School of Physical Science and Technology, Southwest Jiaotong University, Chengdu-610031, PR China}
\email[]{liusm@swjtu.edu.cn}

\footnotetext[1]{Corresponding author: Manpreet Singh}

\begin{abstract}
Pulsar wind electron--positron plasma lacks the heavy inertial species required for the conventional ion-acoustic-like compressive modes. However, if ions are mixed into the reconnecting striped-wind current sheet, a low-frequency compressive branch can appear. We develop a local, comoving-frame theory for such ion-acoustic-like modes in an ion-loaded, pressure-balanced pulsar-wind current sheet.  The background model connects directly to pulsar observables through the light-cylinder magnetic field and Goldreich--Julian density, while the sheet structure is represented by a Harris-like field reversal and species-dependent compression factors.  The pair species are treated as inertialess thermodynamic shielding populations, whereas the ions are described by a warm, nonrelativistic, magnetized fluid.  Pressure balance fixes the pair and ion temperatures rather than prescribing them freely. This gives a closed expression for the pair-shielded ion-acoustic speed in terms of pulsar spin parameters, wind Lorentz factor, pair multiplicity, ion loading fraction, sheet compression, and the ion-to-pair temperature ratio. 
We derive the warm-ion electrostatic dispersion relation and discuss the conditions under which the ion-acoustic-like branch can exist in such a current sheet. We find that the ion-acoustic-like mode does not occur for all sheet parameters but is restricted to specific regions of parameter space. Thus, ion loading alone is not sufficient to sustain the mode; the local current-sheet conditions determine where an admissible ion-acoustic-like mode can exist. Consequently, any wave-driven anomalous dissipation or particle heating mediated by this mode must be highly localized rather than distributed uniformly across the striped wind. This framework provide the physical domain where the mode can exist, providing the necessary foundation for future studies of kinetic excitation and damping.
\end{abstract}

\keywords{pulsar wind---current sheet---electrostatic waves---ion acoustic waves}

\section{Introduction}

Relativistic pulsar winds carry the rotational energy of magnetized neutron stars into their surrounding nebulae, initially in a form dominated by electromagnetic fields rather than particle kinetic energy \citep{KennelCoroniti1984a,Coroniti1990}. Beyond the light cylinder, an oblique rotator produces a striped wind in which toroidal magnetic fields of opposite polarity are separated by thin current sheets \citep{Coroniti1990,LyubarskyKirk2001,KirkSkjaraasen2003,CeruttiBeloborodov2017}. These current sheets are localized sites of magnetic dissipation, where plasma thermal pressure balances the reversing field \citep{Coroniti1990,Kirk2002} and where kinetic instabilities such as the relativistic drift-kink mode can develop \citep{ZenitaniHoshino2005}. The global wind may be described by force-free electrodynamics or ideal magnetohydrodynamics, but these frameworks are insufficient inside the dissipative sheet, where finite particle inertia, thermal pressure, charge separation, and non-ideal electric fields determine the local plasma response.

Although secondary pair cascades can make a pulsar outflow overwhelmingly pair-dominated by number, pair dominance does not require the wind to be baryon-free. On polar-cap field lines for which the Goldreich--Julian corotation charge density is positive, space-charge-limited flow can extract protons or heavier ions from the neutron-star surface, provided that surface binding does not suppress charge emission \citep{GoldreichJulian1969,Jones2012}. Global particle-in-cell calculations that include surface extraction likewise find that energetic ions can enter the open-field-line outflow, escape the magnetosphere, and carry a non-negligible fraction of the outgoing particle-energy flux \citep{PhilippovSpitkovsky2018,GuepinCeruttiKotera2020}. These results provide a physically credible source and transport channel for a dilute ionic component in an otherwise pair-rich pulsar wind.

Whether ions are preferentially concentrated within the equatorial current sheet itself, however, is supported only indirectly. Ion-bearing pulsar-wind and termination-shock models have shown that even a numerically subdominant ion component can influence the shock structure, produce ion-cyclotron fluctuations, transfer energy to the pairs, and reproduce some properties of the Crab wisps \citep{Hoshino1992,GallantArons1994,AmatoArons2006}. Ion-inclusive reconnection simulations further demonstrate that a small ion population, once present in a reconnecting pair-plasma sheet, can participate in plasmoid dynamics and undergo efficient acceleration \citep{ChernoglazovHakobyanPhilippov2023}. These calculations establish the dynamical relevance of ions but do not constitute an observational measurement of the baryon abundance in a striped-wind current sheet. Indeed, global proton-acceleration simulations find that many escaping protons are not confined to the equatorial sheet \citep{GuepinCeruttiKotera2020}. We therefore regard a finite wind ion component as physically motivated, while treating preferential midplane ion enrichment as an explicit and testable model hypothesis rather than an established property of pulsar winds.

In a normal electron–ion plasma, an ion‑acoustic wave (IAW) forms because the heavy ions provide the inertia while the much lighter electrons respond quickly through their thermal pressure \citep{Sagdeev1966,KrallTrivelpiece1973,Gary1993,Stix1992}. Pulsar winds differ fundamentally from this standard electron-ion plasmas because secondary pair cascades produce an outflow dominated by electrons and positrons \citep{GoldreichJulian1969,AronsScharlemann1979,HibschmanArons2001,TimokhinHarding2015}. In a perfectly symmetric electron–positron plasma, both species have the same mass, so there is no light‑versus‑heavy separation to excite a normal acoustic wave. \citep{Tsytovich1970,Verheest2000,MelroseRafatMastrano2021}. Adding even a small number of ions to the plasma introduces the needed heavy inertia, making an acoustic‑like mode possible.

Ion-acoustic (IA) dynamics in electron--positron--ion plasmas have been investigated extensively in homogeneous configurations, exploring the effects of positron concentration, warm ions, magnetic obliquity, and non-thermal particle distributions on linear phase speeds and solitary wave structures \citep{PopelVladimirovShukla1995,Nejoh1996,MahmoodMushtaqSaleem2003,MahmoodAkhtar2008,AlinejadMamun2011,SahaPalChatterjee2014,FerdousiSultanaMamun2015}. More recently, \citet{SinghKakadKakadSaini2021} simulated IA solitary pulses in a weakly relativistic, unmagnetized electron--positron--ion model of a pulsar-wind interaction zone. Nevertheless, existing studies share a fundamental limitation: they treat the plasma as homogeneous with freely prescribed background temperatures and densities. In an active pulsar current sheet, the plasma state cannot be chosen arbitrarily; the thermal pressure, species temperatures, Debye lengths, and gyrofrequencies are strictly constrained by the magnetic-pressure deficit across the field reversal and local current-carrying requirements.

The unresolved question is therefore not whether an idealized electron--positron--ion mixture can support acoustic fluctuations, but whether a self-consistent, pressure-balanced pulsar current sheet admits a physically viable IA-like branch when constrained by observable pulsar parameters. To our knowledge, this work presents the first formulation that embeds a warm-ion electrostatic dispersion relation inside a pressure-balanced Harris-like current sheet connected directly to pulsar spin period ($P$), spin-down rate ($\dot{P}$), wind Lorentz factor ($\Gamma_w$), and pair multiplicity ($\kappa$). We systematically determine the wavenumber window where this mode can propagate by evaluating a comprehensive list of existence criteria, including finite pair shielding, oblique ion magnetization, finite-Larmor-radius validity, current closure, advection residence, and quasi-static sheet evolution.

The paper is organized as follows. Sec.~\ref{sec:model} constructs the pulsar-linked current-sheet equilibrium, including the magnetic field and density profiles, Harris-like field reversal, pressure-balanced temperatures, and pair-shielding scale. Sec.~\ref{sec:fluid} derives the linear warm-ion response and the electrostatic dispersion relation. Sec.~\ref{sec:existence} introduces the conditions for acoustic branch identity, ion magnetization, pair response, ion-fluid validity, local equilibrium, advection, and current closure, and combines them into the allowed wavenumber interval. Sec.~\ref{sec:results} presents the current-sheet profiles and existence domains and examines the dependence of the dispersion on position within the sheet, pair multiplicity, propagation angle, and pulsar $P$--$\dot P$, together with the restoring-force balance. Sec.~\ref{sec:discussion} discusses the physical meaning of these results, their dependence on plasma composition and the host pulsar, and the limits of the fluid description. Sec.~\ref{sec:conclusions} summarizes the main conclusions.

\section{Theoretical Framework}\label{sec:model}

We construct a local model of an ion-loaded pulsar-wind current sheet and use it to determine where an IA-like mode can exist within the sheet. The large-scale pulsar and wind parameters determine the magnetic field, particle densities, and characteristic current-sheet thickness at a chosen radial distance from the neutron star. The local cross-sheet structure is then specified through the magnetic-field reversal, guide field, pair compression, and ion loading. 
Pressure balance then relates the reduction in magnetic pressure inside the sheet to the thermal pressure of the pairs and ions, yielding their local temperature. These quantities define the local ion gyrofrequency, ion plasma frequency, IA speed, and pair-shielding length that enter the wave dispersion relation. The following sections develop each part of this framework, beginning with the large-scale pulsar-wind quantities and then constructing the local current-sheet plasma model used in the wave analysis.

\subsection{Magnetic field model}

We adopt the standard dipole estimate for the surface magnetic field,
\begin{equation}
B_s = 3.2\times10^{19}(P\dot P)^{1/2}\ {\rm [Gauss]},
\label{eq:Bs}
\end{equation}
where $P$ is the pulsar spin period and $\dot{P}$ is its time derivative. The light-cylinder radius is defined as $\RLC=cP/2\pi$
\citep{GoldreichJulian1969}. 

The laboratory frame upstream reconnecting magnetic field at a radial distance $r$ from the neutron star is expressed in terms of the asymptotic toroidal striped‑wind framework \citep{Michel1973,Coroniti1990,SinghKAW2026}, 
\begin{equation}
B_{\rm up}^{\rm lab}(r)=B_s\left(\frac{R_{\rm NS}}{\RLC}\right)^3\frac{\RLC}{r} ,
\label{eq:Bwindlab}
\end{equation}
where $R_{\rm NS}$ is the neutron star radius. Here, the \emph{laboratory frame} denotes the pulsar rest frame, while the \emph{upstream} refers to the uncompressed wind plasma immediately outside the current sheet at a fixed radial distance $r$.

For an ideal-MHD wind, the electric field vanishes in the local fluid comoving frame of the uncompressed plasma outside the current sheet, i.e., $\mathbf E_{\rm up}\simeq 0$.
In the laboratory frame, the same frozen-in condition gives the motional electric field
\begin{equation}
\mathbf E_{\rm up}^{\rm lab}
= -\left(\frac{\mathbf v_w}{c}\right) \times \mathbf B_{\rm up}^{\rm lab},
\label{eq:Elab}
\end{equation}
where $\mathbf v_w$ is the radial bulk velocity of the wind and $\mathbf E_{\rm up}^{\rm lab}$ is the upstream electric field vector measured in the pulsar rest frame \citep[e.g.,][]{Coroniti1990,Kirk2002}. Since the radial wind velocity is orthogonal to the asymptotically toroidal magnetic field, the corresponding field magnitudes satisfy
\begin{equation}
E_{\rm up}^{\rm lab}
=\frac{v_w}{c}
B_{\rm up}^{\rm lab}.
\label{eq:Elab_magnitude}
\end{equation}
In cgs units, the electromagnetic Lorentz invariant gives
\begin{equation}
B_{\rm up}^2-E_{\rm up}^2
=
\left(B_{\rm up}^{\rm lab}\right)^2
-
\left(E_{\rm up}^{\rm lab}\right)^2,
\label{eq:EM_invariant}
\end{equation}
where $E_{\rm up}$ and $B_{\rm up}$ denote the corresponding
comoving-frame field magnitudes. Using $E_{\rm up}\simeq0$ and
Eq.~(\ref{eq:Elab_magnitude}), we obtain
\begin{equation}
B_{\rm up}(r)
=
B_{\rm up}^{\rm lab}(r)
\sqrt{1-\frac{v_w^2}{c^2}}
=
\frac{B_{\rm up}^{\rm lab}(r)}{\Gamma_w},
\label{eq:Bup_comoving}
\end{equation}
where
$\Gamma_w=(1-v_w^2/c^2)^{-1/2}$ is the bulk Lorentz factor.

\subsection{Density model}

The canonical Goldreich--Julian reference density evaluated at the light cylinder is given by 
$n_{{\rm GJ},\rm LC}={B_{\rm LC}}/{e c P}$ \citep{GoldreichJulian1969}, 
where $B_{\rm LC}=B_s(R_{\rm NS}/\RLC)^3$ is the characteristic light-cylinder magnetic field strength. Transforming this scale into the local comoving wind frame accounts for both spherical geometric dilution and relativistic Lorentz contraction, yielding \citep[e.g.,][]{Coroniti1990}:
\begin{equation}
n_{\rm GJ}(r)=\frac{n_{{\rm GJ},\rm LC}}{\Gamma_w}\left(\frac{\RLC}{r}\right)^2.
\label{eq:nGJprime}
\end{equation}

The comoving number densities of the individual plasma species are parameterized across the macroscale current sheet profile as:
\begin{align}
n_{p0}(r,X)&=\kappa n_{\rm GJ}(r)\Cpprof(X),\\
n_{i0}(r,X)&=\eta n_{\rm GJ}(r)\Ciprof(X),\\
n_{e0}(r,X)&=n_{p0}(r,X)+Z_i n_{i0}(r,X),
\label{eq:densitymodel}
\end{align}
where local charge neutrality is strictly imposed. Here, $\kappa$ denotes the secondary pair multiplicity factor originating from electrostatically or curvature-driven pair cascades \citep{TimokhinHarding2015,TimokhinHarding2019}, while $\eta$ scales the baseline ion loading (ion contamination via surface extraction or entrainment) in units of the comoving Goldreich--Julian density \citep[see][]{Hoshino1992}. Spatial localized compression within the current layer is governed by the species-dependent profiles:
\begin{equation}
\mathcal{C}_q(X)=1+A_q\sech^2\left(\frac{X}{\Deltap}\right),
\label{eq:Cprofiles}
\end{equation}
where $q=p\,{(\rm positrons)},i \,{(\rm ions)}$;  $A_q$ represent the dimensionless maximum compression amplitudes for the positrons and ions, and $\Delta$ represents the current sheet thickness. 
We assume preferential ion enrichment near the sheet center, such that $A_i>A_p$. Existing studies support the presence of ions in pulsar outflows and demonstrate their dynamical importance once they enter a reconnecting current layer, but they do not presently constrain the relative pair and ion compression factors in a striped-wind current sheet \citep{GuepinCeruttiKotera2020,ChernoglazovHakobyanPhilippov2023}. We therefore treat $A_i>A_p$ as a model assumption and examine how such localized ion enrichment affects the IA-like mode.

To diagnose the local composition and track whether the sheet core remains pair-dominated or undergoes significant ion modification, we define the spatial charge ratio:
\begin{equation}
\alpha(r,X)=\frac{n_{p0}(r,X)}{Z_i n_{i0}(r,X)}=\frac{\kappa \,\Cpprof(X)}{Z_i\eta\,\Ciprof(X)} .
\label{eq:alpha}
\end{equation}
This ratio provides a direct measure of the relative pair shielding versus ion loading within the sheet. 
Values of $\alpha(r,X) \gg 1$ correspond to a pair‑dominated regime where electron–positron contributions outweigh ions, while $\alpha(r,X) \lesssim 1$ represents a significant ion loading.

\subsection{Harris-like current sheet}

At a fixed radial distance $(r)$, the reconnecting magnetic field is modeled as a local Harris reversal \citep{Harris1962}:
\begin{equation}
B_{Y0}(r,X)=B_{\rm up}(r)\tanh\left(\frac{X}{\Delta}\right),
\label{eq:HarrisB}
\end{equation}
where $X$ denotes the macroscopic coordinate across the current sheet.
This is supplemented by a uniform guide field component:
\begin{equation}
B_{g0}(r)=\varepsilon_gB_{\rm up}(r) ,
\label{eq:guide}
\end{equation}
where $\varepsilon_g$ denotes the ratio of the guide field $B_{g0}(r)$ to the reconnecting magnetic field $B_{\rm up}(r)$, thereby quantifying the relative strength of the out-of-plane magnetic component.
Thus, the total local magnetic field magnitude is given by:
\begin{equation}
B_{\rm loc,0}(r,X)=B_{\rm up}(r) \left[\tanh^2\left(\frac{X}{\Deltap}\right)+\varepsilon_g^2\right]^{1/2}.
\label{eq:Bloc}
\end{equation}
The sustaining equilibrium current density follows directly from Ampere's law:
\begin{equation}
J_{\rm req}(r,X)=\frac{c}{4\pi}\frac{\partial B_{Y0}}{\partial X} =\frac{cB_{\rm up}(r)}{4\pi\Delta}\sech^2\left(\frac{X}{\Delta}\right).
\label{eq:Jreq}
\end{equation}
Equations~(\ref{eq:HarrisB})--(\ref{eq:Jreq}) represent a local magnetostatic equilibrium. This treatment is valid under the assumption that the wave period $\omega^{-1}$ (where $\omega$ is the wave frequency) is short compared to the characteristic macroscale reconnection evolution time $\tau_{\rm rec}$, satisfying the timescale separation $\omega\tau_{\rm rec}\gg1$.

Rather than allowing the sheet half-thickness to vary with the local, $X$-dependent compressed density, we tie the half-thickness to a fixed, upstream total pair inertial length scale:
\begin{equation}
\Delta(r)=\zeta d_{\ell,{\rm up}}(r),
\label{eq:Delta}
\end{equation}
where $\zeta$ is a constant dimensionless scaling factor, and the upstream total-pair inertial length is written as:
\begin{equation}
 d_{\ell,{\rm up}}(r)=c\left[\frac{m_e h_{\ell,{\rm up}}} {4\pi e^2(2\kappa+Z_i\eta)\, n_{\rm GJ}(r)}\right]^{1/2}.
\label{eq:dskin}
\end{equation}
Here $h_{\ell,{\rm up}}$ is the dimensionless upstream total pair enthalpy factor, included because $d_{\ell,{\rm up}}$ is a dynamical inertial scale. 

\subsection{Pressure-balanced temperature}

The temperature profile across the sheet is set by local pressure balance, ensuring a self‑consistent thermal structure
\begin{equation}
P_{{\rm tot},0}(r,X)
+ \frac{B_{\rm loc,0}^{\,2}(r,X)}{8\pi}
=
P_{{\rm up},0}(r)+ \frac{B_{\rm up}^{2}(r)(1+\varepsilon_g^2)}{8\pi}.
\label{eq:pressurebalance}
\end{equation}
We define the upstream asymptotic plasma pressure relative to the magnetic pressure via the upstream plasma beta parameter, $\beta_{\rm up}$, such that \citep{Kirk2002},
\begin{equation}
P_{\rm up,0}(r)=\beta_{\rm up}\frac{B_{\rm up}^2(r)}{8\pi}.
\end{equation}
By substituting the magnetic field profile from Eq.~(\ref{eq:Bloc}), the spatial distribution of the total thermal plasma pressure across the sheet reduces to:
\begin{equation}
P_{\rm tot,0}(r,X)=\frac{B_{\rm up}^2(r)}{8\pi} \left[\beta_{\rm up}+\sech^2\left(\frac{X}{\Delta}\right)\right].
\label{eq:Ptot}
\end{equation}

To partition this pressure among the individual components, we assume a common thermodynamic temperature $T_{\ell0}$ for the pair species ($T_{e0} = T_{p0} = T_{\ell0}$) and parameterize the thermal ion-to-pair ratio via the partition variable $\tau_i \equiv T_{i0}/T_{\ell0}$. The total multi-species thermal pressure is then governed by the ideal gas law
\begin{equation}
\begin{split}
P_{{\rm tot},0}(r,X) =
\big[n_{e0}(r,X)+n_{p0}(r,X)\big]k_BT_{\ell0}(r,X)\\+
n_{i0}(r,X)k_BT_{i0}(r,X).
\label{eq:Ptot_species}
\end{split}
\end{equation}
Combining this framework with the individual species density profiles yields a closed, analytical expression for the local pair temperature across the layer
\begin{equation}
k_BT_{\ell0}(r,X) = \frac{B_{\rm up}^{2}(r)S(X)}{8\pi
\left[2\kappa\mathcal{C}_p(X)+(Z_i+\tau_i)\eta\mathcal{C}_i(X)
\right] n_{\rm GJ}(r)},
\label{eq:Tpair}
\end{equation}
where the localized pressure modulation factor is defined as
\begin{equation}
S(X)\equiv \beta_{\rm up}+\sech^2\left(\frac{X}{\Delta}\right),
\end{equation}
and the localized warm-ion temperature profile is fixed by:
\begin{equation}
T_{i0}(r,X) = \tau_iT_{\ell0}(r,X).
\label{eq:Ti}
\end{equation}

Equation~(\ref{eq:Tpair}) determines the local temperature from pressure balance between the magnetic and thermal components of the current sheet. As the reversing magnetic field decreases toward the sheet center, the associated loss of magnetic pressure is balanced by an increase in thermal pressure. The resulting temperature depends on both this magnetic-pressure decrease and the local pair and ion densities.

\subsection{Pair shielding closure}
\label{sub:pair-shielding}

In the present model, the electron and positron populations contribute to the electrostatic response through their local static density response. In the nondegenerate classical limit, the equilibrium density of species $s$ can be written in the general form
\begin{equation}
n_s(T_s,\mu_s)
=
F_s(T_s)
\exp\left(\frac{\mu_s}{k_B T_s}\right),
\qquad s=e,p,
\label{eq:classical_density_response}
\end{equation}
where $\mu_s$ is the chemical potential and $F_s(T_s)$ contains the temperature-dependent part of the equilibrium distribution. At fixed temperature, Eq.~(\ref{eq:classical_density_response}) gives
\begin{equation}
\chi_s
\equiv
\left.
\left(
\frac{\partial n_s}{\partial\mu_s}
\right)_{T_s}
\right|_0
=
\frac{n_{s0}}{k_B T_{s0}},
\qquad
\label{eq:susceptibility}
\end{equation}
where the subscript $0$ denotes evaluation at the local current-sheet equilibrium. This relation holds for both Maxwell--Boltzmann and Maxwell--J\"uttner distributions in the nondegenerate classical limit, since their different momentum dependences are contained in $F_s(T_s)$ \citep{Synge1957}.

The electrons and positrons both contribute to electrostatic shielding. Their combined response therefore gives
\begin{equation}
\lambda_{\rm pair}^{-2}(r,X)
=
4\pi e^2
\left[
\frac{n_{e0}(r,X)}{k_B T_{e0}(r,X)}
+
\frac{n_{p0}(r,X)}{k_B T_{p0}(r,X)}
\right].
\label{eq:lambda_pair_general}
\end{equation}
For the common local pair temperature adopted here, $T_{e0}=T_{p0}\equiv T_{\ell0}$, this becomes
\begin{equation}
\lambda_{\rm pair}^{-2}(r,X)
=
\frac{
4\pi e^2
\left[
n_{e0}(r,X)+n_{p0}(r,X)
\right]
}{
k_B T_{\ell0}(r,X)
}.
\label{eq:lambda_pair}
\end{equation}
Thus, the pair-shielding length varies across the current sheet through both the local pair densities and the temperature obtained from the pressure-balanced sheet model.

No explicit pair enthalpy factor appears in Eq.~(\ref{eq:lambda_pair}) because the shielding length is determined by the static thermodynamic density response. In the nondegenerate classical limit, this response remains $\chi_s=n_{s0}/(k_B T_{s0})$ even when the equilibrium momentum distribution is relativistic. The pair enthalpy factor $h_{\ell,{\rm up}}$ instead enters the dynamical pair inertia and hence the total-pair inertial length $d_{\ell,{\rm up}}$ used to parameterize the current-sheet thickness $\Delta$. It therefore does not enter $\lambda_{\rm pair}$ explicitly, although it can influence the local wave properties indirectly through the adopted current-sheet structure.

\section{Derivation of dispersion relation}\label{sec:fluid}

In the low‑frequency IA-like compressive mode studied in this paper, inertia is supplied entirely by the heavy ions. We therefore derive the warm, magnetized ion‑fluid response from first principles, keeping the treatment local. Electrons and positrons, being much lighter, respond almost instantaneously to the wave and are represented by static thermodynamic susceptibilities rather than separate dynamical momentum equations. As a result, the only time‑dependent momentum dynamics we follow explicitly are those of the three‑component ion fluid.

\subsection{Governing Fluid Equations}

We work in a local comoving frame centered at a fixed macroscopic position $(r,X)$ inside the current sheet. The equilibrium magnetic field is treated as locally uniform over the shorter wave scale and is written as
\begin{equation}
\mathbf B_0 = B_0\hat{\mathbf z},
\qquad B_0\equiv B_{\rm loc,0}(r,X).
\label{eq:B0_local}
\end{equation}
High‑frequency wave perturbations act on much smaller scales and are described using the local coordinates $\mathbf{x}=(x,y,z)$, whereas the equilibrium quantities vary slowly on the macroscopic coordinate $X$. For compactness, the explicit dependence of the equilibrium quantities on $(r,X)$ is dropped throughout the local derivation below, except where 
$r$ and $X$ must appear explicitly.
Thus, $n_{(e,p,i)0}$, $T_{(e,p,i)0}$, $B_0$, $\Omega_i$, $c_{s,i}$, $\lambda_{\rm pair}$, and $\omega_{pi}$ denote their local values evaluated at the selected macroscopic position.

The warm ion population is described by its comoving number density $n_i$, charge $Z_i e$, mass $m_i$, and three-component fluid velocity vector $\mathbf{u}_i = (u_x, u_y, u_z)$. Under a localized electrostatic perturbation characterized by the potential $\phi$, the high-frequency electric field reduces to $\mathbf{E} = -\nabla\phi$. The ion continuity and momentum equations are written respectively as
\begin{equation}
\frac{\partial n_i}{\partial t}+\nabla\cdot(n_i\mathbf{u}_i)=0,
\label{eq:ion_cont_exact}
\end{equation}
\begin{equation}
\begin{split}
m_i n_i\left(\frac{\partial}{\partial t}+\mathbf{u}_i\cdot\nabla\right)\mathbf{u}_i &= Z_i e n_i\left(-\nabla\phi+\frac{\mathbf{u}_i}{c}\times\mathbf{B}_0\right)\\ 
&\quad-\nabla P_i.
\label{eq:ion_mom_exact}
\end{split}
\end{equation}
Expanding the momentum equation into its component forms along the wave coordinates $(x, y, z)$ yields:
\begin{equation}
\begin{split}
m_i n_i\left(\frac{\partial u_x}{\partial t} + u_x\frac{\partial u_x}{\partial x} + u_y\frac{\partial u_x}{\partial y} + u_z\frac{\partial u_x}{\partial z}\right) \\= Z_i e n_i\left(-\frac{\partial \phi}{\partial x} + \frac{u_y B_{0}}{c}\right) - \frac{\partial P_i}{\partial x},
\label{eq:mom_x_comp}
\end{split}
\end{equation}

\begin{equation}
\begin{split}
m_i n_i\left(\frac{\partial u_y}{\partial t} + u_x\frac{\partial u_y}{\partial x} + u_y\frac{\partial u_y}{\partial y} + u_z\frac{\partial u_y}{\partial z}\right) \\= Z_i e n_i\left(-\frac{\partial \phi}{\partial y} - \frac{u_x B_{0}}{c}\right) - \frac{\partial P_i}{\partial y},
\label{eq:mom_y_comp}
\end{split}
\end{equation}

\begin{equation}
\begin{split}
m_i n_i\left(\frac{\partial u_z}{\partial t} + u_x\frac{\partial u_z}{\partial x} + u_y\frac{\partial u_z}{\partial y} + u_z\frac{\partial u_z}{\partial z}\right) \\= Z_i e n_i\left(-\frac{\partial \phi}{\partial z}\right) - \frac{\partial P_i}{\partial z}.
\label{eq:mom_z_comp}
\end{split}
\end{equation}

\noindent
To close the ion fluid equations, we assume an isotropic polytropic pressure response. At each fixed macroscopic position $(r,X)$, we define the local ion-pressure response speed by
\begin{equation}
c_{s,i}^{2}
\equiv
\frac{1}{m_i}
\left(\frac{\partial P_i}{\partial n_i}
\right)_0 ,
\label{eq:csi_general}
\end{equation}
where the derivative is evaluated about the local current-sheet equilibrium. For the polytropic closure
\(P_i\propto n_i^{\gamma_i}\), this coefficient becomes
\begin{equation}
c_{s,i}^{2}
=
\frac{\gamma_i k_B T_{i0}}{m_i},
\label{eq:csi_speed_def}
\end{equation}
where $\gamma_i$ is the ion adiabatic index and $T_{i0}$ is determined by the pressure-balanced equilibrium in Eq.~(\ref{eq:Ti}).

The highly mobile electrons and positrons are assumed to satisfy a local Maxwell-Boltzmann distribution function which gives following densities for electrons and positrons,  respectively, 
\begin{equation}
n_e = n_{e0}
\exp\left(\frac{e\phi}{k_BT_{e0}}
\right), \quad n_p = n_{p0} \exp\left( -\frac{e\phi}{k_B T_{p0}} \right).
\label{eq:ne_np_boltz}
\end{equation}
The electrostatic potential profile is self-consistently tied to the species distribution via Poisson's equation, which can be written as:
\begin{equation}
\nabla^2\phi = -4\pi e \left( Z_i n_i + n_p - n_e \right).
\label{eq:poisson_exact}
\end{equation}
At unperturbed magnetostatic equilibrium, the background species densities naturally satisfy local charge neutrality: $Z_i n_{i0}(r,X) + n_{p0}(r,X) = n_{e0}(r,X)$.

\subsection{Linearization}

We perturb the system about the local equilibrium with small‑amplitude fluctuations: 
\begin{equation}
\begin{aligned}
n_i &= n_{i0}+n_{i1},\quad n_e = n_{e0}+n_{e1},\\
n_p &= n_{p0}+n_{p1}, \quad P_i = P_{i0}+P_{i1},\\
\mathbf{u}_i &= \mathbf{u}_{i1},\quad \phi = \phi_1, 
\end{aligned}
\end{equation}
where the terms with `1' in subscripts denotes the first‑order (small) perturbations about the equilibrium.
Thus, linearizing the ion continuity equation (\ref{eq:ion_cont_exact}) yields:
\begin{equation}
\frac{\partial n_{i1}}{\partial t} + n_{i0} \left( \frac{\partial u_{x1}}{\partial x} + \frac{\partial u_{y1}}{\partial y} + \frac{\partial u_{z1}}{\partial z} \right) = 0.
\label{eq:ion_cont_lin}
\end{equation}
Similarly, linearizing the momentum balance Eqs.~(\ref{eq:mom_x_comp})--(\ref{eq:mom_z_comp}) and introducing the localized ion gyrofrequency $\Omi \equiv Z_i e B_{0}/ (m_i c)$,  we obtain, respectively

\begin{align}
\frac{\partial u_{x1}}{\partial t} &= -\frac{Z_i e}{m_i}\frac{\partial \phi_1}{\partial x} + \Omi u_{y1} - \frac{c_{s,i}^2}{n_{i0}}\frac{\partial n_{i1}}{\partial x}, \label{eq:lin_mom_x} \\
\frac{\partial u_{y1}}{\partial t} &= -\frac{Z_i e}{m_i}\frac{\partial \phi_1}{\partial y} - \Omi u_{x1} - \frac{c_{s,i}^2}{n_{i0}}\frac{\partial n_{i1}}{\partial y}, \label{eq:lin_mom_y} \\
\frac{\partial u_{z1}}{\partial t} &= -\frac{Z_i e}{m_i}\frac{\partial \phi_1}{\partial z} - \frac{c_{s,i}^2}{n_{i0}}\frac{\partial n_{i1}}{\partial z}. \label{eq:lin_mom_z}
\end{align}

Expansion of exponential terms in Eqs.~(\ref{eq:ne_np_boltz}) to first order yields the linear pair density responses:
\begin{equation}
n_{e1} \approx n_{e0} \frac{e\phi_1}{k_B T_{e0}}, \qquad n_{p1} \approx -n_{p0} \frac{e\phi_1}{k_B T_{p0}}.
\label{eq:lin_pair_responses}
\end{equation}
Substituting these expressions into the linearized Poisson's Eq.~(\ref{eq:poisson_exact}), gives
\begin{equation}
\nabla^2\phi_1 = -4\pi e \left[ Z_i n_{i1} - e\phi_1\left( \frac{n_{e0}}{k_B T_{e0}} + \frac{n_{p0}}{k_B T_{p0}} \right)\right]. \nonumber
\label{eq:poisson_intermediate}
\end{equation}
Recalling the definition of the combined thermodynamic pair shielding length scale from Eq.~(\ref{eq:lambda_pair_general}), Poisson's equation simplifies to:
\begin{equation}
\nabla^2\phi_1 - \lambda_{\rm pair}^{-2}\phi_1 = -4\pi Z_i e n_{i1}.
\label{eq:poisson_final_lin}
\end{equation}

\subsection{Fourier Analysis and Dielectric Response}

We assume a plane‑wave solution of the form $\exp[i(\mathbf{k}\cdot\mathbf{x}-\omega t)]$ for the linearized Eqs.~(\ref{eq:ion_cont_lin})-(\ref{eq:poisson_final_lin}), which results in $\partial / \partial t \rightarrow -i\omega$ and $\nabla \rightarrow i\mathbf{k}$. Without loss of generality we orient the wavevector in the $x$--$z$ plane, $\mathbf{k}=(k_\perp,0,k_\parallel)$, so that $k^2=k_\perp^2+k_\parallel^2$.

Defining the normalized ion density perturbation variable $N_1 \equiv n_{i1} / n_{i0}$, the Fourier-transformed continuity equation becomes:
\begin{equation}
\omega N_1 = k_\perp u_{x1} + k_\parallel u_{z1} .
\label{eq:fourier_cont}
\end{equation}
To write the linearized momentum equations compactly, we introduce the generalized first-order potential 
\begin{equation}
\Psi_1
\equiv
\frac{Z_i e\phi_1}{m_i}
+
c_{s,i}^{2}N_1 .
\label{eq:Psi1}
\end{equation}

\noindent
Applying this transformation, the linearized momentum Eqs.~(\ref{eq:lin_mom_x})--(\ref{eq:lin_mom_z}) becomes
\begin{align}
-i\omega u_{x1} &= -ik_\perp \Psi_1 + \Omi u_{y1} , \label{eq:fourier_mom_x} \\
-i\omega u_{y1} &= -\Omi u_{x1} , \label{eq:fourier_mom_y} \\
-i\omega u_{z1} &= -ik_\parallel \Psi_1 . \label{eq:fourier_mom_z}
\end{align}

\noindent
Solving the transverse $y$-component yields $u_{y1} = -i (\Omi / \omega) u_{x1}$. Substituting this back into Eq.~(\ref{eq:fourier_mom_x}) allow us to isolate the cross-field velocity
\begin{equation}
u_{x1} = \frac{\omega k_\perp}{\omega^2 - \Omi^2} \Psi_1.
\label{eq:ux_fourier_sol}
\end{equation}
Simplification of Eq.~(\ref{eq:fourier_mom_z}) yields parallel velocity component
\begin{equation}
u_{z1} = \frac{k_\parallel}{\omega} \Psi_1 .
\label{eq:uz_fourier_sol}
\end{equation}
Substitution of localized velocity solutions (\ref{eq:ux_fourier_sol}) and (\ref{eq:uz_fourier_sol}) back into the continuity relation (\ref{eq:fourier_cont}) gives
\begin{equation}
N_1 = \left( \frac{k_\perp^2}{\omega^2 - \Omi^2} + \frac{k_\parallel^2}{\omega^2} \right) \Psi_1.
\label{eq:N_intermediate_fourier}
\end{equation}
Grouping the structural geometric projection terms into a single dielectric scalar factor
\begin{equation}
Q(\omega, \mathbf{k}) \equiv \frac{k_\perp^2}{\omega^2 - \Omi^2} + \frac{k_\parallel^2}{\omega^2} ,
\label{eq:A_factor_def}
\end{equation}
Eq.~(\ref{eq:N_intermediate_fourier}) simplifies to $N_1 = Q \Psi_1$. Expanding $\Psi_1$ back into its constituent terms via Eq.~(\ref{eq:Psi1}) yields
\begin{equation}
N_1 = Q \left( \frac{Z_i e \phi_1}{m_i} + c_{s,i}^2 N_1 \right) \implies N_1 = \frac{Q}{1 - Q c_{s,i}^2}\frac{Z_i e \phi_1}{m_i} .
\label{eq:N_final_fourier}
\end{equation}
Multiplying by $n_{i0}$ on both sides, we arrive at the exact warm magnetized ion fluid density response
\begin{equation}
n_{i1} = \frac{Z_i e n_{i0}}{m_i} \frac{Q}{1 - Q c_{s,i}^2} \phi_1 .
\label{eq:ni1_final_response}
\end{equation}

\subsection{Dispersion Relation}
\label{sec:dispersion}

With the analytical fluid response derived, we now construct the full dispersion relation without invoking cold-ion approximation, or low-frequency ($\omega \ll \Omi$) limitation.
Applying the spatial Fourier transform ($\nabla^2 \rightarrow -k^2$) to the linearized multi-species Poisson equation (\ref{eq:poisson_final_lin}),  yields
\begin{equation}
\left( k^2 + \lambda_{\rm pair}^{-2} \right) \phi_1 = 4\pi Z_i e n_{i1}.
\label{eq:poisson_fourier}
\end{equation}
Substituting the ion fluid density response from Eq.~(\ref{eq:ni1_final_response}) into Eq.~(\ref{eq:poisson_fourier}) yields the expression
\begin{equation}
\left( k^2 + \lambda_{\rm pair}^{-2} \right) \phi_1 = \left( \frac{4\pi Z_i^2 e^2 n_{i0}}{m_i} \right) \frac{Q}{1 - Q c_{s,i}^2} \phi_1 .
\label{eq:disp_cancel_step}
\end{equation}
Defining the local ion plasma frequency as $\omega_{pi}^2 \equiv 4\pi Z_i^2 e^2 n_{i0} / m_i$ and canceling the non-trivial electrostatic potential ($\phi_1 \neq 0$), yields the fundamental implicit dispersion relation
\begin{equation}
\left( k^2 + \lambda_{\rm pair}^{-2} \right) \left( 1 - Q c_{s,i}^2 \right) = \omega_{pi}^2 Q,
\label{eq:implicit_dispersion}
\end{equation}
where $Q(\omega, \mathbf{k})$ remains explicitly defined by Eq.~(\ref{eq:A_factor_def}). Expanding this expression out fully yields the general, unreduced electrostatic fluid dispersion relation
\begin{equation}
\begin{split}
\left( k^2 + \lambda_{\rm pair}^{-2} \right) \left[ 1 - c_{s,i}^2 \left( \frac{k_\perp^2}{\omega^2 - \Omi^2} + \frac{k_\parallel^2}{\omega^2} \right) \right] \\= \omega_{pi}^2 \left( \frac{k_\perp^2}{\omega^2 - \Omi^2} + \frac{k_\parallel^2}{\omega^2} \right).
\label{eq:full_explicit_dispersion}
\end{split}
\end{equation}

To isolate the individual physical propagating modes, we reconstruct the dispersion framework into an explicit quartic polynomial form. Let $K \equiv k^2 + \lambda_{\rm pair}^{-2}$ describe the collective short-wavelength shielding filter. Rearranging Eq.~(\ref{eq:implicit_dispersion}) gives
\begin{equation}
K = Q \left( \omega_{pi}^2 + K c_{s,i}^2 \right) .
\label{eq:K_rearrange}
\end{equation}
By finding a common denominator for the components of the dielectric projection operator $Q$, we can rewrite Eq.~(\ref{eq:A_factor_def}) as
\begin{equation}
Q = \frac{k_\perp^2 \omega^2 + k_\parallel^2(\omega^2 - \Omi^2)}{\omega^2 (\omega^2 - \Omi^2)} = \frac{k^2 \omega^2 - k_\parallel^2 \Omi^2}{\omega^2 (\omega^2 - \Omi^2)} .
\label{eq:A_polynomial_form}
\end{equation}
Substituting this form back into Eq.~(\ref{eq:K_rearrange}) and multiplying both sides by the denominator isolates the spectral terms
\begin{equation}
K \omega^2 \left( \omega^2 - \Omi^2 \right) = \left( \omega_{pi}^2 + K c_{s,i}^2 \right) \left( k^2 \omega^2 - k_\parallel^2 \Omi^2 \right) .
\label{eq:polynomial_pre_divide}
\end{equation}
Dividing the entire expression by $K$,  yields
\begin{equation}
\omega^2 \left( \omega^2 - \Omi^2 \right) = \left( c_{s,i}^2 + \frac{\omega_{pi}^2}{k^2 + \lambda_{\rm pair}^{-2}} \right) \left( k^2 \omega^2 - k_\parallel^2 \Omi^2 \right) .
\label{eq:polynomial_divided}
\end{equation}
Now, we define the spatial effective multi-species acoustic speed profile $C_{\rm IA}$, which physically accounts for the shared warm-ion restoration pressure and the screening susceptibility of the surrounding pair background
\begin{equation}
C_{\rm IA}^{\,2}(k;r,X)\equiv c_{s,i}^2 + \frac{\omega_{pi}^2}{k^2 + \lambda_{\rm pair}^{-2}}.
\label{eq:Cs_effective_def}
\end{equation}
Applying this compact formulation, Eq.~ (\ref{eq:polynomial_divided}) collapses into a standard quartic polynomial equation in terms of the wave frequency
\begin{equation}
\omega^4 - \omega^2 \left( \Omi^2 + k^2 C_{\rm IA}^2\right) + k_\parallel^2 \Omi^2 C_{\rm IA}^2 = 0.
\label{eq:quartic_final_dispersion}
\end{equation}
The quadratic formula in $\omega^2$ then yields the two roots
\begin{multline}
\omega^2 = \frac{1}{2} \bigg[ \left(\Omi^2 + k^2 C_{\rm IA}^2\right)\\ \pm \sqrt{ \left(\Omi^2 + k^2 C_{\rm IA}^2\right)^2 - 4 k_\parallel^2 \Omi^2 C_{\rm IA}^2} \,\bigg] .
\label{eq:quartic_roots}
\end{multline}
The upper root $\omega_{\rm IC}$ with ($+$) sign gives the high-frequency IC-like branch, in which cross-field ion gyromotion controls the response. The lower root $\omega_{\rm IA}$ with ($-$) sign gives the low-frequency IA-like branch. This second branch is the mode of interest here, and it serves as the conditional diagnostic of local ion loading examined in the remainder of the paper. For parallel propagation ($k_\perp = 0$, $k_\parallel = k$), the gyromotion decouples from the compressive dynamics and the quartic factors as $(\omega^2 - \Omi^2)(\omega^2 - k^2 C_{\rm IA}^2) = 0$. 
The acoustic branch then reduces to the familiar unmagnetized form $\omega_{\rm ac}^2 = k^2 C_{\rm IA}^2$.
At oblique angles this decoupling fails: the magnetic field ties the perpendicular ion velocity to the wave, the phase speed acquires an explicit dependence on $\Omi$ and the propagation angle, and the full three-component fluid treatment must be retained.

\section{existence and Local Existence Domain}
\label{sec:existence}

The two solutions of Eq.~(\ref{eq:quartic_roots}) are formal roots of the warm magnetized fluid system.  The lower root can be identified as an IA-like mode only over the range of wavenumbers for which its frequency remains acoustic-like and the assumptions used in the local fluid model remain valid. 
Below, we discuss the conditions that determine whether the lower solution, $\omega_{\rm IA}(k;r,X)$, belongs to the pair-shielded, magnetized-ion, fluid regime considered here.

At each macroscopic position $(r,X)$, the admissible wavenumber range is bounded at high $k$ by the change in branch identity, finite ion magnetization, pair-response limits, and finite-ion-gyroradius effects. At low $k$, it is bounded by the requirement that the wave evolve sufficiently rapidly compared with the current-sheet evolution and advection timescales, while remaining sufficiently local relative to the spatial variation of the equilibrium. Current closure and the nonrelativistic ion-temperature condition impose additional position-dependent constraints.

For compactness, we introduce
\begin{equation}
\xi\equiv\frac{X}{\Delta},
\qquad
k_\parallel=k\cos\theta,
\qquad
k_\perp=k\sin\theta ,
\label{eq:existence_geometry}
\end{equation}
and evaluate all equilibrium quantities at the selected position
$(r,X)$.

\subsection{Acoustic identity of the lower root}
\label{subsec:acoustic_identity}

For parallel propagation, Eq.~(\ref{eq:quartic_final_dispersion})
factorizes exactly as
\begin{equation}
\left(\omega^2-\Omega_i^2\right)
\left(\omega^2-k^2C_{\rm IA}^2\right)=0.
\label{eq:parallel_factorization_adm}
\end{equation}
The two uncoupled solutions are therefore
\begin{equation}
\omega_{\rm ac}^2=k^2C_{\rm IA}^2,
\qquad
\omega_{\rm ci}^2=\Omega_i^2.
\label{eq:parallel_reference_branches}
\end{equation}
They exchange their frequency ordering at the transition wavenumber
$k_{\rm tr}$ defined by
\begin{equation}
k_{\rm tr}C_{\rm IA}(k_{\rm tr};r,X)
=
\Omega_i(r,X).
\label{eq:ktr_existence}
\end{equation}
For $k<k_{\rm tr}$, the lower solution is the IA solution, whereas for $k>k_{\rm tr}$ the lower solution follows the
IC branch.

At finite obliquity, the two responses are coupled and the exact crossing is replaced by a smooth transition. The long-wavelength acoustic reference frequency is then
\begin{equation}
\omega_{\rm ac,0}^2(k;r,X)
\equiv
k_\parallel^2 C_{\rm IA}^2(k;r,X).
\label{eq:oblique_acoustic_reference}
\end{equation}
We measure the acoustic character of the lower root using
\begin{equation}
f_{\rm ac}(k;r,X)
=
\frac{
\left|\omega_{\rm IA}^2-\Omega_i^2\right|
}{
\left|\omega_{\rm IA}^2-\omega_{\rm ac,0}^2\right|
+
\left|\omega_{\rm IA}^2-\Omega_i^2\right|
}.
\label{eq:fac_definition}
\end{equation}
This quantity approaches unity when the lower root is closer to the acoustic reference and approaches zero when it is closer to the IC frequency.  We classify the lower root as predominantly acoustic when $f_{\rm ac}(k;r,X)\geq f_{\rm ac,min}$ and $f_{\rm ac,min}=0.5$.
The corresponding upper boundary $k_{\rm ac}(r,X)$ is defined by $f_{\rm ac}(k_{\rm ac};r,X)=f_{\rm ac,min}$. Thus, the acoustic-identity condition normally admits $k\leq k_{\rm ac}(r,X)$. 
The value $f_{\rm ac,min}=0.5$ is a branch-classification threshold: at this value the lower root is equally separated from the acoustic and IC reference frequencies.

\subsection{Ion-magnetization limit}
\label{subsec:ion_magnetization}

Although the exact quartic retains the full dependence on $\Omega_i$, an IA interpretation requires the wave
frequency to remain well below the local ion gyrofrequency.  We write
this condition as
\begin{equation}
\frac{\omega_{\rm IA}(k;r,X)}{\Omega_i(r,X)}
\leq
\epsilon_\Omega ,
\label{eq:magnetization_condition}
\end{equation}
where the fiducial value is $\epsilon_\Omega=0.1$.
The corresponding upper wavenumber is obtained from $\omega_{\rm IA}(k_\Omega;r,X)
=\epsilon_\Omega\Omega_i(r,X)$. 
The admitted range is therefore $k\leq k_\Omega(r,X)$. This condition becomes most important where the local magnetic field is weakest, e.g., the smallest value of $k_\Omega$ generally occurs near the current-sheet center. The guide field prevents $\Omega_i$ from vanishing there,
but the ion-magnetization interval can still be substantially narrower
than in the outer sheet.

\subsection{Pair-response limits}
\label{subsec:pair_existence}

The electron and positron populations are described through their local thermodynamic density response. For this approximation to remain valid, two conditions must be satisfied: the wavelength must be sufficiently large compared with the local pair-shielding scale, and the pair populations must respond on a timescale shorter than the wave period.

The pair-shielding condition is written as $k\lambda_{\rm pair}(r,X) \leq \epsilon_\lambda$ and $\epsilon_\lambda=1$. 
It gives the explicit upper boundary for wavenumber
\begin{equation}
k_\lambda
=
\frac{\epsilon_\lambda}
{\lambda_{\rm pair}(r,X)}.
\label{eq:klambda_boundary}
\end{equation}

To determine whether pair inertia can be neglected, we introduce the local total-pair response frequency
\begin{equation}
\omega_\ell^2(r,X)
=
\frac{
4\pi e^2
\left[n_{e0}(r,X)+n_{p0}(r,X)\right]
}{
m_e h_{\ell,0}(r,X)
}.
\label{eq:omega_lepton_local}
\end{equation}
Here $h_{\ell,0}$ is the local dimensionless pair enthalpy factor.
In the present numerical calculation it is approximated by the upstream value $h_{\ell,{\rm up}}$.  The temporal pair-response
condition is
\begin{equation}
\frac{\omega_{\rm IA}(k;r,X)}
{\omega_\ell(r,X)}
\leq
\epsilon_\ell,
\qquad
\epsilon_\ell=0.1.
\label{eq:pair_temporal_condition}
\end{equation}
Its upper boundary $k_\ell(r,X)$ satisfies $\omega_{\rm IA}(k_\ell;r,X) =
\epsilon_\ell\omega_\ell(r,X)$.

The pair-response interval is consequently, $k\leq k_{\rm pair}(r,X)$ and $k_{\rm pair} \equiv \min\left(k_\lambda,k_\ell\right)$. 
The two parts have different meanings: $k\lambda_{\rm pair}\leq1$ restricts the wavelength relative to the shielding scale, while $\omega_{\rm IA}/\omega_\ell\leq0.1$ checks whether the pair populations can follow the wave without requiring their inertia to be evolved
explicitly.

\subsection{Ion-fluid limits}
\label{subsec:ion_fluid_existence}

The ion momentum equations do not include finite-Larmor-radius or
gyroviscous corrections.  The perpendicular wavelength must therefore
remain larger than the ion gyroradius.  With
\begin{equation}
v_{{\rm th},i}(r,X)
=
\left(\frac{k_BT_{i0}(r,X)}{m_i}\right)^{1/2}
=
\frac{c_{s,i}(r,X)}{\sqrt{\gamma_i}},
\label{eq:vthi_adm}
\end{equation}
the local ion gyroradius is
\begin{equation}
\rho_i(r,X)
=
\frac{v_{{\rm th},i}(r,X)}
{\Omega_i(r,X)}
=
\frac{c_{s,i}(r,X)}
{\sqrt{\gamma_i}\,\Omega_i(r,X)}.
\label{eq:rhoi_adm}
\end{equation}
We impose
\begin{equation}
k_\perp\rho_i
=
k\sin\theta\,\rho_i
\leq
\epsilon_{\rm FLR},
\qquad
\epsilon_{\rm FLR}=0.1.
\label{eq:FLR_condition}
\end{equation}
For $\theta\neq0$, this gives
\begin{equation}
k_{\rm FLR}(r,X)
=
\frac{\epsilon_{\rm FLR}}
{\rho_i(r,X)\sin\theta}.
\label{eq:kFLR_boundary}
\end{equation}

The nonrelativistic ion momentum equation also requires
\begin{equation}
\Theta_i(r,X)
\equiv
\frac{k_BT_{i0}(r,X)}{m_i c^2}
=
\frac{c_{s,i}^2(r,X)}
{\gamma_i c^2}
\leq
\Theta_{i,\max},
\label{eq:Theta_i_condition}
\end{equation}
for which we adopt $\Theta_{i,\max}=0.03$. 
This is a position-dependent condition and does not itself define a
wavenumber boundary.

Finally, the characteristic acoustic speed and lower-root phase speed
must remain subluminal:
\begin{equation}
\frac{C_{\rm IA}(k;r,X)}{c}<1,
\qquad
\frac{\omega_{\rm IA}(k;r,X)}{kc}<1.
\label{eq:subluminal_conditions}
\end{equation}
The complete ion-fluid domain is the intersection of
Eqs.~(\ref{eq:FLR_condition}), (\ref{eq:Theta_i_condition}), and
(\ref{eq:subluminal_conditions}).

\subsection{Quasi-static and local-equilibrium limits}
\label{subsec:local_equilibrium_existence}

The dispersion relation is evaluated at a fixed macroscopic position
inside a slowly evolving current sheet.  The wave must therefore
complete several oscillations before the background changes
appreciably.  We parameterize the sheet-evolution time as
\begin{equation}
\tau_{\rm rec}
=
\frac{L_{\rm rec}}{v_{\rm rec}}
=
\frac{\chi_{\rm rec}\Delta}
{\epsilon_{\rm rec}c},
\label{eq:tau_rec_definition}
\end{equation}
where $L_{\rm rec} \equiv \chi_{\rm rec}\Delta$ and $ v_{\rm rec} \equiv
\epsilon_{\rm rec}c$.
Here $L_{\rm rec}$ is the assumed macroscopic length of the evolving
current-sheet segment, generally measured along the sheet.  It is not
the cross-sheet thickness, which is of order $2\Delta$ for the
Harris profile.

The number of wave cycles completed during $\tau_{\rm rec}$ is
\begin{equation}
N_{\rm qs}(k;r,X)
=
\frac{\omega_{\rm IA}(k;r,X)\tau_{\rm rec}}{2\pi}.
\label{eq:Nqs_definition}
\end{equation}
We require $N_{\rm qs}(k;r,X)\geq
N_{\rm qs,min}$ and $N_{\rm qs,min}=3.$
The corresponding lower wavenumber $k_{\rm qs}(r,X)$ is defined by
\begin{equation}
\omega_{\rm IA}(k_{\rm qs};r,X)
=
\frac{2\pi N_{\rm qs,min}}{\tau_{\rm rec}}.
\label{eq:kqs_boundary}
\end{equation}
The fiducial calculation adopts $\chi_{\rm rec}=100$ and $\epsilon_{\rm rec}=0.1$, so that $\tau_{\rm rec}=1000\Delta/c$.  These values specify a macroscopic scale-separation model and are not universal properties of a pulsar current sheet.

The local derivation also assumes that the equilibrium coefficients vary weakly over one wavelength in the direction in which the wave samples the cross-sheet structure.  Let $\varphi_k$ denote the angle between $\mathbf{k}_\perp$ and the macroscopic sheet-normal direction $\hat{\mathbf X}$. The sheet-normal wavevector component is $k_X = k\,g_X$, where $g_X \equiv \left|\sin\theta\cos\varphi_k\right|.$
For each local coefficient
\begin{equation}
a(r,X)
\in
\mathcal{M}
\equiv
\left\{
\Omega_i,\,
\omega_{pi},\,
c_{s,i},\,
\lambda_{\rm pair}
\right\},
\label{eq:inhomogeneous_coefficients}
\end{equation}
we define the first- and second-order inhomogeneity measures
\begin{equation}
\epsilon_{1,a} = \frac{\left|\partial_X\ln a\right|}{k_X},\qquad \epsilon_{2,a} =\frac{\left|\partial_X^2\ln a\right|}{k_X^2}.
\label{eq:inhomogeneity_measures}
\end{equation}
The combined local-inhomogeneity parameter is
\begin{equation}
\epsilon_{\rm inh}(k;r,X)
=
\max_{q\in\mathcal{M}}
\left(
\epsilon_{1,a},
\epsilon_{2,a}
\right).
\label{eq:epsilon_inh_definition}
\end{equation}
The second-order term is retained because the first derivative vanishes at symmetry points such as the sheet center even when the equilibrium has finite curvature.

We require $\epsilon_{\rm inh}(k;r,X) \leq
\epsilon_{\rm inh,max}$, where $\epsilon_{\rm inh,max}=0.1$.
For $g_X\neq0$, this gives the explicit lower boundary
\begin{equation}
\begin{split}
k_{\rm inh}(r,X)
= \max_{q\in\mathcal{Q}}
\bigg[
&
\frac{\left|\partial_X\ln a\right|}{\epsilon_{\rm inh,max}g_X},
\\
&
\frac{1}{g_X} \left(\frac{\left|\partial_X^2\ln a\right|}{\epsilon_{\rm inh,max}}\right)^{1/2}
\bigg].
\label{eq:kinh_boundary}
\end{split}
\end{equation}
The local-equilibrium domain is consequently, $k\geq k_{\rm local}(r,X)$ and $k_{\rm local} \equiv
\max\left(k_{\rm qs},k_{\rm inh}\right)$.
When $k_X=0$, the wave does not sample the cross-sheet gradient through its plane-wave phase, and the local approximation must instead be specified through a finite wave-packet width.

\subsection{Advection residence time}
\label{subsec:advection_existence}

A mode that satisfies the local dispersion relation must also complete
at least one oscillation before leaving the selected current-sheet
pocket.  We write the residence time as
\begin{equation}
t_{\rm adv}
=
\frac{L_\parallel}{v_{\rm flow}}
=
\frac{\chi_\parallel\Delta}{v_{\rm flow}},
\label{eq:tadv_definition}
\end{equation}
where $L_\parallel=\chi_\parallel\Delta$ is the longitudinal size of
the local pocket.  The number of completed cycles is
\begin{equation}
N_{\rm adv}(k;r,X) = \frac{\omega_{\rm IA}(k;r,X)t_{\rm adv}}{2\pi}.
\label{eq:Nadv_definition}
\end{equation}
We impose $N_{\rm adv}(k;r,X) \geq
N_{\rm adv,min}$ and $
N_{\rm adv,min}=1$. 
The lower boundary is therefore determined by
\begin{equation}
\omega_{\rm IA}(k_{\rm adv};r,X)
=
\frac{2\pi N_{\rm adv,min}}{t_{\rm adv}}.
\label{eq:kadv_boundary}
\end{equation}
The fiducial calculation adopts $\chi_\parallel=1000$ and $v_{\rm flow}=c$.

\subsection{Current closure}
\label{subsec:current_closure_existence}

The magnetic-field reversal across the current sheet requires a current that can be carried by the local plasma. The required current density is given by Eq.~(\ref{eq:Jreq}). The maximum current that can be provided by the available electrons, positrons, and ions is estimated as
\begin{equation}
J_{\rm av}(r,X) = ec \left[n_{e0}(r,X) + n_{p0}(r,X) + Z_i n_{i0}(r,X) \right].
\label{eq:Javailable}
\end{equation}
We compare the required and available currents through
\begin{equation}
\Xi_J(r,X)
\equiv
\frac{J_{\rm req}(r,X)}
{J_{\rm av}(r,X)},
\label{eq:XiJ_definition}
\end{equation}
so that the current-sheet profile can be supported when $\Xi_J(r,X)<1$.
Here, $J_{\rm av}$ corresponds to the maximum current obtained if the charge carriers drift at speeds up to $c$. It therefore provides a simple upper limit on the current that the local plasma can carry, rather than specifying the drift velocity of each species. This condition varies with position across the current sheet through the local densities and required current, but it does not depend on the wave number $k$.

\subsection{Combined admissible wavenumber interval}
\label{subsec:combined_existence}

The conditions above define a lower and an upper wavenumber boundary
at each position $(r,X)$. The lower boundary is set by the
quasi-static, local-inhomogeneity, and advection conditions,
\begin{equation}
k_{\rm low}(r,X)
=
\max
\left[
k_{\rm qs}(r,X),
k_{\rm inh}(r,X),
k_{\rm adv}(r,X)
\right].
\label{eq:klow_definition}
\end{equation}
The upper boundary is determined by the acoustic-identity,
ion-magnetization, pair-response, and finite-Larmor-radius limits,
\begin{equation}
\begin{aligned}
k_{\rm high}(r,X)
= \min\!\big[
& k_{\rm ac}(r,X),\; k_{\Omega}(r,X), \\
& k_{\rm pair}(r,X),\; k_{\rm FLR}(r,X)
\big].
\end{aligned}
\label{eq:khigh_definition}
\end{equation}
An IA-like fluid mode can be identified only where
\begin{equation}
k_{\rm low}(r,X)
<
k
<
k_{\rm high}(r,X),
\label{eq:admissible_interval}
\end{equation}
provided that the ion-temperature, subluminal-speed, and
current-closure conditions are also satisfied. These limits provide
the local wavenumber range used to interpret the existence maps
below.

\section{Results and Interpretation}\label{sec:results}

We evaluate the pressure-balanced fluid model using a fiducial set of parameters representative of a young energetic pulsar such as the Crab \citep{Kirk2002}. The adopted values are $P=0.033,{\rm s}$, $\dot P=4.21\times10^{-13},{\rm s\,s^{-1}}$, $\Gamma_w=250$, $r=50R_{\rm LC}$, $\kappa=10^5$, $\eta=10$, $A_p=10$, $A_i=10^2$, $\varepsilon_g=0.1$, $\tau_i=0.1$, $\gamma_i=5/3$, $\beta_{\rm up}=10^{-7}$, $h_{\ell,{\rm up}}$ and $\zeta=100$. The choice $\Gamma_w=250$ and the radial range $r=10-100R_{\rm LC}$ are guided by relativistic striped-wind models of the Crab pulsar \citep{Kirk2002}. The adopted pair multiplicity lies within the broad range allowed by pulsar pair-cascade calculations \citep{TimokhinHarding2015,TimokhinHarding2019}. In contrast, the ion loading and the stronger compression assigned to the ions are not observationally constrained. We therefore treat them as conditional model parameters and use them to determine under what current-sheet conditions an IA-like branch can exist.

\begin{figure*}[t]
\centering
\includegraphics[width=01.95\columnwidth]{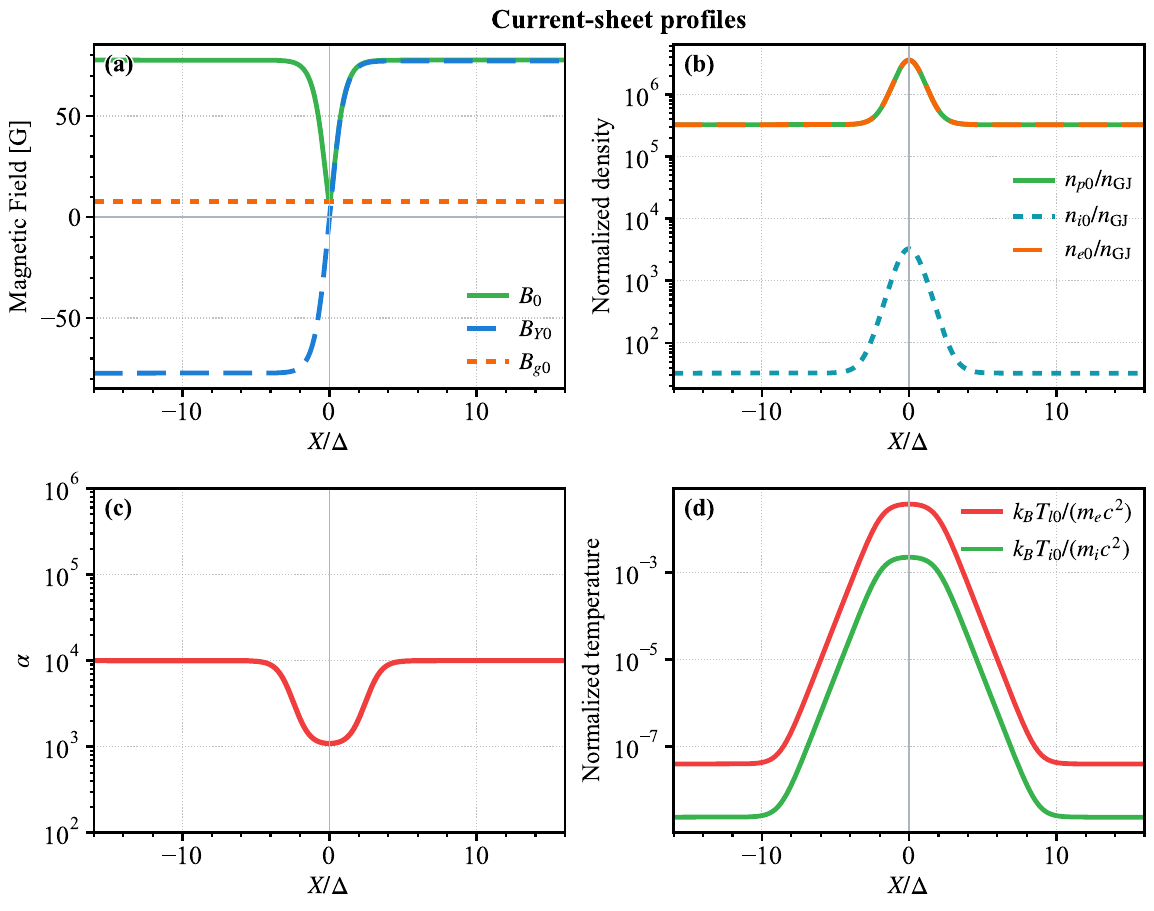}
\caption{Equilibrium plasma profiles across the pressure-balanced current sheet at $r=50R_{\rm LC}$. Panel (a) shows the reversing field $B_{Y0}$, guide field $B_{g0}$, and total local field $B_0\equiv B_{\rm loc,0}$. Panel (b) shows the electron, positron, and ion densities normalized by $n_{\rm GJ}$, while panel (c) gives the local pair-to-ion charge ratio $\alpha=n_{p0}/(Z_i n_{i0})$. Panel (d) shows the pressure-balanced pair and ion temperatures, normalized as $k_B T_{\ell0}/(m_e c^2)$ and $k_B T_{i0}/(m_i c^2)$, respectively.}
\label{fig:field-density-temp.}
\end{figure*}

\subsection{Current Sheet Profiles}

\subsubsection{Field Configuration}

Figure~\ref{fig:field-density-temp.}(a) illustrate the macroscopic comoving magnetic field topology across the current sheet as a function of the normalized cross-sheet coordinate $X/\Delta$. The blue-dashed curve illustrates the cross-sheet transition of the reversing toroidal component, $B_{Y0}/B_{\rm up}$, which follows a traditional hyperbolic tangent trajectory. This component drops smoothly from its highly magnetized asymptotic upstream values, passing through zero at the exact sheet center ($X=0$) to establish a classic magnetic null geometry. 
We additionally include a uniform out-of-plane guide field, $B_{g0}=\varepsilon_g B_{\rm up}$, shown by the orange dashed curve. This component does not produce the Harris pressure balance, but it prevents the total equilibrium field magnitude from vanishing at the
sheet center.
As a result, the total local magnetic-field magnitude, $B_0$, decreases toward the center of the current sheet, reaches a finite minimum at $X=0$, and increases again toward its upstream value on either side.
This finite base magnetic field fulfill a primary physical requirement for the low-frequency magnetized configuration developed in this work. If the guide field ratio is set to zero ($\varepsilon_g = 0$), the local magnetic field would vanish entirely from the center of the current sheet. 
Since the local ion cyclotron frequency is proportional to the local field magnitude ($\Omega_i \propto B_{\rm 0}$), an unguided current sheet implies $\Omega_i \rightarrow 0$ at $X=0$.
Such a singular breakdown would make the magnetized fluid approximation invalid at the middle of the current sheet, where the prescribed density compression is strongest and where the local dispersion is evaluated.
In the present model, the guide field limits the perpendicular ion response and preserves a finite local ion gyrofrequency at the sheet center.
This localized baseline magnetic field establishes a nonzero minimum field strength, which preserves the local cyclotron frequency for oblique wave vectors and thus supports the validity of our linear fluid expansion near center of the current sheet.

\subsubsection{Density compression and local composition}
\label{subsec:density_results}

Figures~\ref{fig:field-density-temp.}(b) and ~\ref{fig:field-density-temp.}(c) shows the pair and ion density profiles, and positron-to-ion charge ratio ($\alpha$), respectively.
As shown in Fig.~\ref{fig:field-density-temp.}(b), within the current sheet, the pair density follows $\mathcal{C}_p(X)=1+A_p\sech^2(X/\Delta)$, while the ion density follows $\mathcal{C}_i(X)=1+A_i\sech^2(X/\Delta)$.
Because the fiducial model adopts $A_i=10^3$ and $A_p=10$, the ions are compressed much more strongly than the secondary pairs near the sheet center. Both populations return to their upstream values outside the current sheet, where $\mathcal{C}_p,\mathcal{C}_i\rightarrow1$.

As shown in Fig.~\ref{fig:field-density-temp.}(c), far from the sheet ($|X|/\Delta\gg1$), $\alpha\simeq10^4$ for the adopted singly charged ions, confirming that the upstream plasma is strongly pair dominated. 
Near the sheet center ($X/\Delta=0$), the stronger ion compression reduces the ratio to $\alpha\simeq10^3$. Despite this reduction, the positron density remains substantially larger than the ion charge density, and the pair population continues to provide the dominant electrostatic shielding and charge-balancing response. The reduced value of $\alpha$, however, makes the relative contribution of ion inertia less strongly suppressed in the central region.
This reduction in $\alpha$ is an important conditional ingredient of the model. If $A_i$ were comparable to $A_p$, the ratio would remain close to its upstream value. Ion inertia would still be present, but its influence on the electrostatic response would be comparatively weak because of the much larger pair density. Preferential ion compression reduces this imbalance and allows the ion component to contribute more appreciably to the local wave dynamics.

The adopted density profile allows for localized ion enrichment within the current sheet. Relativistic reconnection simulations show that current layers can fragment into magnetic islands or plasmoids that contain compressed plasma and energetic particles \citep{Sironi2014}. However, the preferential accumulation of ions relative to pairs has not been established for the pulsar-wind conditions considered here. We therefore treat the stronger ion compression as a conditional model assumption rather than as a direct prediction of existing kinetic simulations.
Although the wind remains pair dominated overall, a locally enhanced ion fraction within the sheet core reduces the imbalance between ion inertia and the pair shielding response. Under these conditions, the ion contribution to the local electrostatic response becomes more significant, although the existence of the mode must still satisfy the shielding, magnetization, and current-closure constraints derived earlier.

\subsubsection{Pressure-Balanced Temperature Profiles}

Figure~\ref{fig:field-density-temp.}(d) shows the equilibrium pair and ion temperature profiles, expressed as $k_BT_{\ell0}/(m_ec^2)$ and $k_BT_{i0}/(m_ec^2)$, respectively.
These temperatures are not prescribed independently. Instead, they follow from the transverse pressure-balance condition together with the assumed density profiles. Both temperatures increase symmetrically toward the current-sheet center and reach a broad maximum near $X=0$. The ion temperature has the same spatial dependence as the pair temperature but remains lower by the fixed factor $T_{i0}/T_{\ell0}=\tau_i$.

The central enhancement results from the balance between two competing effects. As the reversing magnetic field decreases toward the sheet center, the thermal pressure must increase to maintain the total pressure across the sheet. 
At the same time, the pair and ion densities also increase because of the prescribed compression. The local temperature is determined by the ratio of the required thermal pressure to the corresponding density-weighted particle content. 
For the adopted parameters, the increase in thermal pressure is stronger than the increase in this density factor, producing the temperature maximum seen in Fig.~\ref{fig:field-density-temp.}(d). The density compression partially limits the temperature increase and accounts for the broad shape of the central peak, but it is not strong enough to produce a central depression.

This closure ties the thermal state of the plasma directly to the current-sheet equilibrium and avoids treating the temperatures as unrelated free parameters. The enhanced ion temperature near the sheet center increases the local ion thermal speed, $c_{s,i}^2$. The pair temperature also modifies the shielding length through $\lambda_{\rm pair}^{-2}\propto [n_{e0}+n_{p0}]/T_{\ell0}$. Consequently, the local IA-like speed $C_{\rm IA}$ is controlled jointly by the temperature and density profiles rather than by a simple proportionality to $T_{\ell0}^{1/2}$.

\subsection{existence domains across the current sheet}
\label{subsec:existence_results}

\begin{figure*}[t]
\centering
\includegraphics[width=1.0\textwidth]{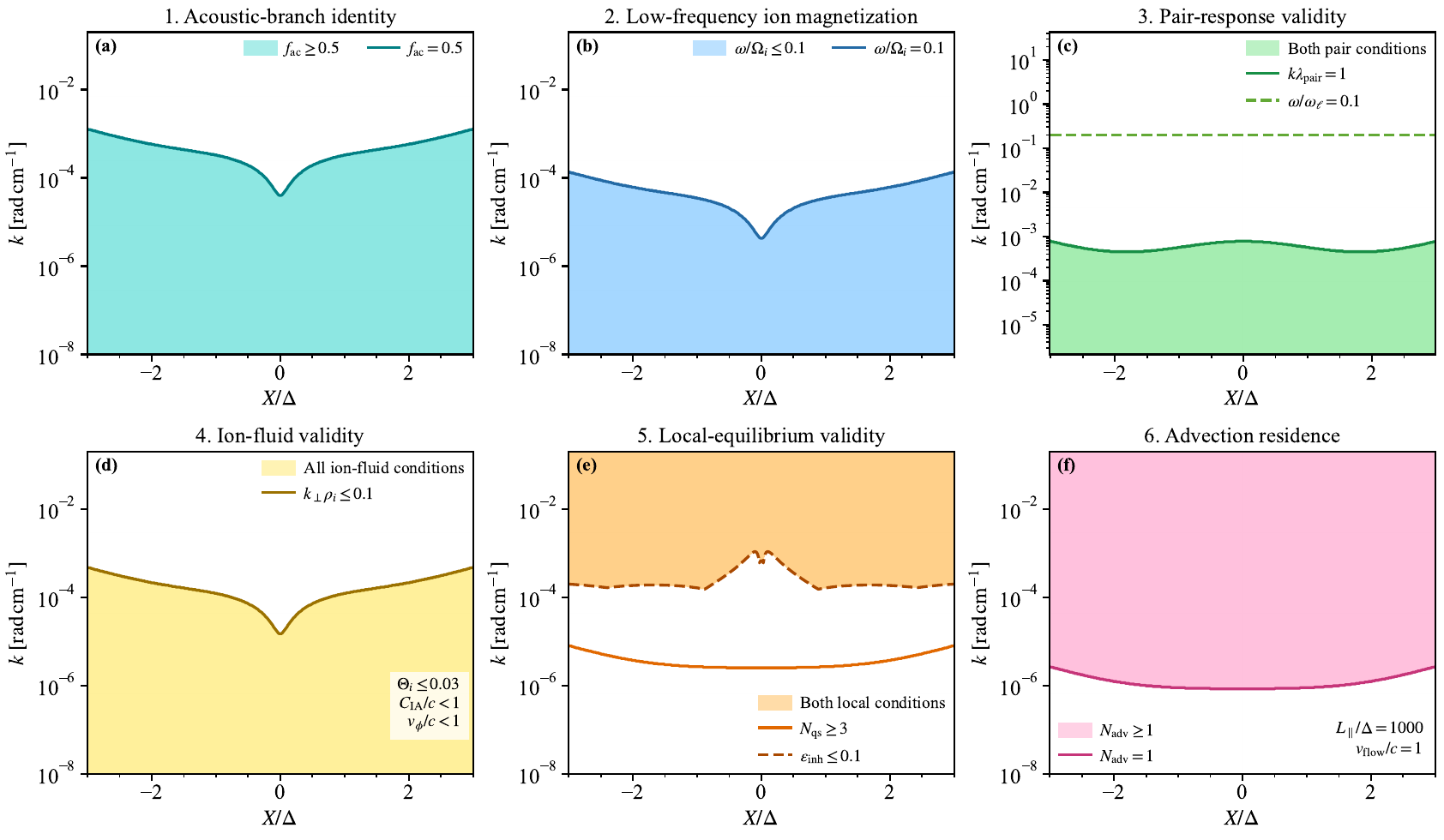}
\caption{Local existence domains of the lower electrostatic branch across the current sheet. Shaded regions show where each condition is satisfied, and the curves mark the corresponding boundary values. Panel (a) shows the acoustic-identity condition $f_{\rm ac}\geq0.5$, and panel (b) the ion-magnetization condition $\omega_{\rm IA}/\Omega_i\leq0.1$. Panel (c) combines the spatial and temporal pair-response conditions, $k\lambda_{\rm pair}\leq1$ and $\omega_{\rm IA}/\omega_\ell\leq0.1$. Panel (d) shows the combined ion-fluid conditions, panel (e) the quasi-static and local-inhomogeneity conditions, and panel (f) the advection condition $N_{\rm adv}\geq1$. The wave geometry is  $\theta=20^\circ$ and $\varphi_k=0^\circ$.}
\label{fig:existence}
\end{figure*}

Figure~\ref{fig:existence} shows how the different validity conditions restrict the IA-like branch across the current sheet. Panels~(a)--(d) mainly set upper limits on the allowed wavenumber, whereas panels~(e) and~(f) set lower limits. The upper limits arise when increasing $k$ moves the solution away from the acoustic regime or brings it into a range where ion magnetization, pair response, or the ion-fluid approximation is no longer adequate. The lower limits arise because sufficiently long-wavelength waves evolve too slowly compared with the sheet evolution or advection time, or sample too much of the spatial variation of the background during one wavelength.

\begin{figure}[t]
\centering
\includegraphics[width=0.90\columnwidth]{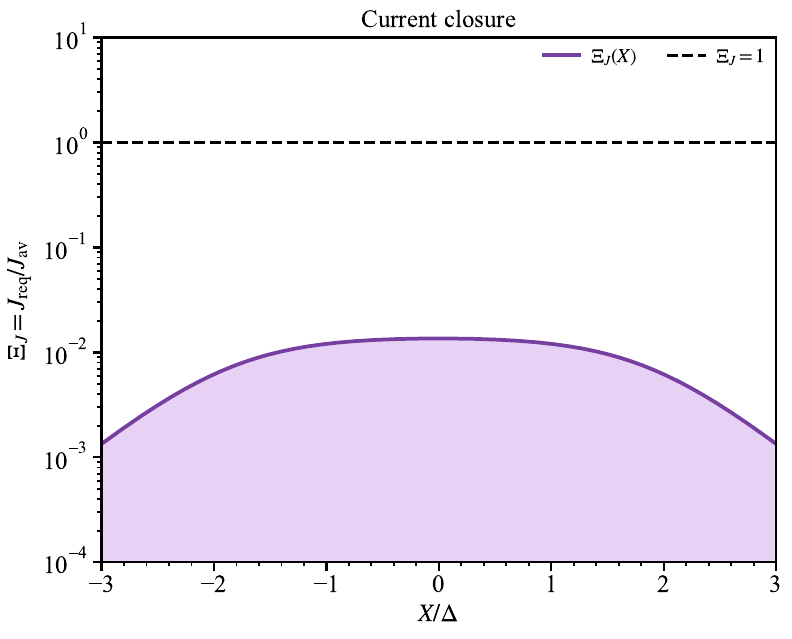}
\caption{Current closure across the pressure-balanced current sheet. The solid curve shows $\Xi_J=J_{\rm req}/J_{\rm av}$, and the horizontal dashed line marks the limiting value $\Xi_J=1$. The condition $\Xi_J<1$ is satisfied throughout the plotted region for the fiducial parameters, with the largest current demand occurring near the sheet center.}
\label{fig:current-closure}
\end{figure}

Panel~(a) shows the acoustic identity of the lower branch. The colored region satisfies $f_{\rm ac}\geq0.5$, and its upper edge defines $k_{\rm ac}(r,X)$. The allowed range narrows toward the sheet center, where the local magnetic field and ion gyrofrequency are smallest. The transition from an IA-like response toward the IC response therefore occurs at a smaller wavenumber near $X=0$ than in the outer part of the sheet.

Panel~(b) applies the low-frequency ion-magnetization condition ${\omega_{\rm IA}}/{\Omega_i}\leq0.1$. 
The corresponding boundary $k_{\Omega}(r,X)$ also reaches its minimum near the sheet center because $\Omega_i$ follows the local magnetic field strength. The finite guide field keeps $\Omega_i$ nonzero at $X=0$, but the central region still provides the strongest restriction on the low-frequency magnetized-ion response.

Panel~(c) shows the two limits associated with the pair response. The solid curve marks the spatial shielding boundary, $k\lambda_{\rm pair}=1$, while the dashed curve marks the temporal response boundary, $\omega_{\rm IA}/\omega_\ell=0.1$. The spatial boundary varies across the current sheet through the local pair-shielding length and occurs at substantially smaller wavenumbers than the temporal boundary. In contrast, the temporal boundary lies at much larger $k$ and changes only weakly across the sheet. The allowed pair-response region must satisfy both conditions and is therefore controlled mainly by the spatial shielding limit for the fiducial parameters considered here.

Panel~(d) shows the ion-fluid domain obtained from the combined requirements:  $k_{\perp}\rho_i\leq0.1$, $\Theta_i\leq0.03$, ${C_{\rm IA}}/{c}<1$ and ${\omega_{\rm IA}}/{kc}<1$.  
For the fiducial parameters, the ion temperature and the two speed conditions remain within their adopted limits throughout the plotted region. The visible upper boundary is therefore controlled mainly by the finite-ion-gyroradius condition. It moves to smaller $k$ toward the sheet center because the weaker magnetic field reduces $\Omega_i$ and increases the local ion gyroradius $\rho_i$.

Panel~(e) shows the two lower-wavenumber limits associated with the local current-sheet treatment. The solid curve is defined by $N_{\rm qs} =3$, while the dashed curve corresponds to $\epsilon_{\rm inh}=0.1$. The allowed region lies above both curves, so the effective local-equilibrium boundary is $k_{\rm local}=\max(k_{\rm qs},k_{\rm inh})$. For the adopted parameters, the inhomogeneity condition is more restrictive over most of the sheet. Its pronounced structure near the sheet center results from the curvature of the equilibrium profiles: the first spatial derivatives vanish at $X=0$ by symmetry, while the second derivatives remain finite.

Panel~(f) gives the advection limit. The boundary is defined by $N_{\rm adv}=1$, so that the colored region contains waves that complete at least one oscillation during the adopted residence time. For $L_{\parallel}=1000\Delta$ and $v_{\rm flow}=c$, the resulting $k_{\rm adv}$ lies below the local inhomogeneity boundary over most of the sheet. Advection therefore provides a weaker lower-wavenumber restriction than the local-equilibrium condition for this fiducial case.

Figure~\ref{fig:current-closure} shows the current-closure ratio $\Xi_J=J_{\rm req}/J_{\rm av}$ across the current sheet. Unlike the conditions shown in Fig.~\ref{fig:existence}, $\Xi_J$ depends only on the position $X/\Delta$ and does not introduce a separate wavenumber boundary. The ratio increases toward the sheet center, where the current required to support the magnetic-field reversal is largest, and decreases toward the outer parts of the sheet. For the fiducial parameters, its maximum value remains of order $10^{-2}$, well below the condition $\Xi_J=1$.

Thus, the available charge carriers are sufficient to support the required Harris-sheet current throughout the plotted region. The current-closure condition therefore does not further reduce the allowed IA-like wavenumber interval for the fiducial case. Here, $J_{\rm av}$ represents the maximum current that the local electrons, positrons, and ions could carry if their drift speeds approached $c$; the condition $\Xi_J<1$ therefore checks the available current-carrying capacity rather than specifying the actual drift velocity of each species.

The individual wavenumber limits shown in Fig.~\ref{fig:existence} combine through Eqs.~(\ref{eq:klow_definition}) and (\ref{eq:khigh_definition}). At each position, the IA-like fluid branch is allowed only when $k_{\rm low}<k<k_{\rm high}$. The current-closure condition is shown separately in Fig.~\ref{fig:current-closure} because it depends on position but not on $k$. Together, Fig.~\ref{fig:existence} and
\ref{fig:current-closure} give the  wavenumber-dependent and position-dependent conditions used to define the local IA-like domain.

\begin{figure*}[t]
\centering
\includegraphics[width=1.95\columnwidth]{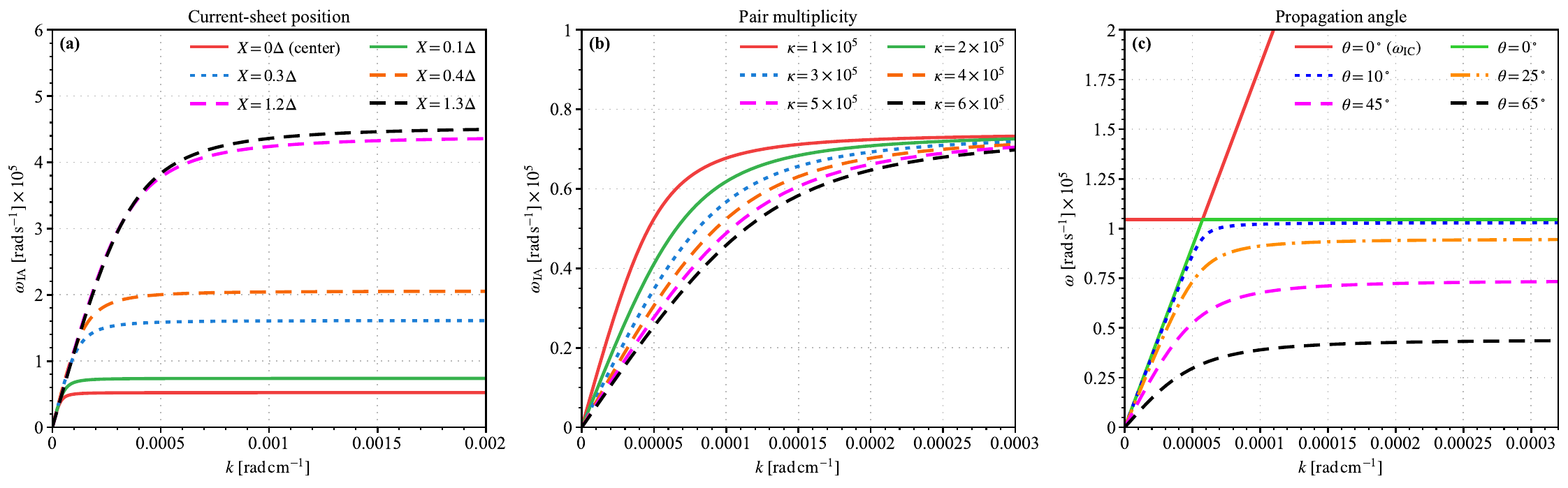}
\caption{Dispersion of the electrostatic branches for different current-sheet and plasma parameters. Panel (a) shows the lower-root frequency $\omega_{\rm IA}(k)$ at $X/\Delta=0$, $0.1$, $0.3$, $0.4$, $1.2$, and $1.3$. Panel (b) shows the effect of pair multiplicity, $\kappa=(1$--$6)\times10^5$, at the sheet center. Panel (c) shows the angular dependence from $\theta=0^\circ$ to $65^\circ$; the upper IC branch $\omega_{\rm IC}$ for $\theta=0^\circ$ is also shown to identify the exact acoustic--IC crossing in parallel propagation. All other parameters are held at their fiducial values. Panels (a) and (b) use $\theta=20^\circ$.}
\label{fig:frequency-vs-k}
\end{figure*}

\begin{figure*}[t]
\centering
\includegraphics[width=1.0\textwidth]{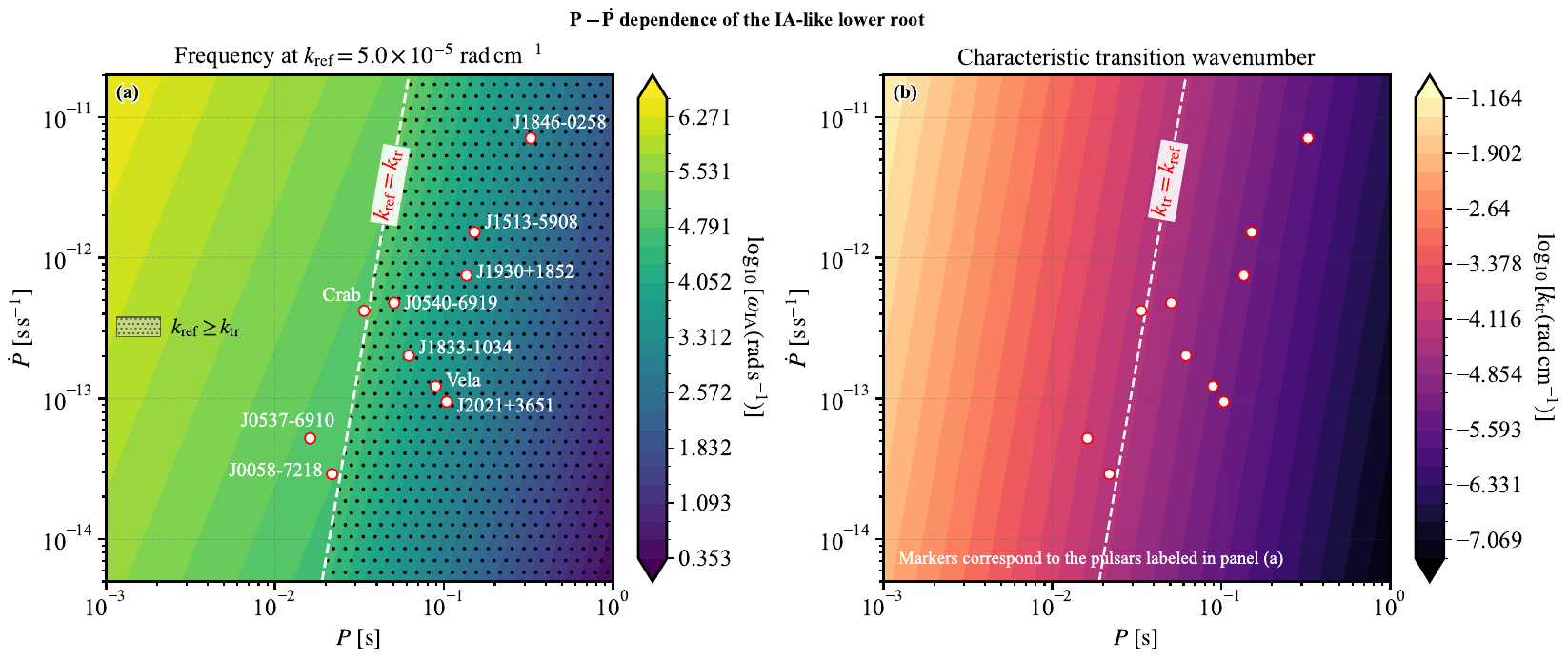}
\caption{$P$--$\dot P$ dependence of the lower electrostatic branch for $r=50R_{\rm LC}$, $X/\Delta=0.1$, and $\theta=0^\circ$. Panel (a) shows the lower-root frequency at the fixed wavenumber $k_{\rm ref}=5.0\times10^{-5}\ {\rm rad\,cm^{-1}}$, while panel (b) shows the acoustic--IC transition wavenumber $k_{\rm tr}$ defined by $k_{\rm tr}C_{\rm IA}(k_{\rm tr})=\Omega_i$. The dashed contour marks $k_{\rm ref}=k_{\rm tr}$. In panel (a), the dotted region has $k_{\rm ref}\geq k_{\rm tr}$, where the lower root follows the IC solution; outside this region it follows the acoustic solution. Marked sources indicate the measured $P$ and $\dot P$ of selected young energetic pulsars, with all other current-sheet parameters held fixed.}
\label{fig:P-Pdot-map}
\end{figure*}

\subsection{Dispersion Characteristics}

\subsubsection{Spatial Variation of the Wave Dispersion across the Current Sheet}
\label{subsec:spatial-wave-dispersion}

Figure~\ref{fig:frequency-vs-k}(a) shows $\omega_{\rm IA}$ as a function of $k$ at six positions across the current sheet: $X/\Delta=0$, $0.1$, $0.3$, $0.4$, $1.2$, and $1.3$. The lowest frequencies are found at the sheet center, while the curves move progressively upward away from the center. The largest frequencies occur at $X/\Delta=1.2$ and $1.3$, where the local magnetic field has recovered most of its upstream strength.

At small $k$, the lower branch follows the nearly linear IA-like relation $\omega_{\rm IA}\simeq k_\parallel C_{\rm IA}$. The difference between the curves in this range therefore comes mainly from the change of $C_{\rm IA}$ across the sheet, since the local density, temperature, ion plasma frequency, and pair-shielding length all vary with $X$. As $k$ increases, ion gyromotion becomes more important. The branch then departs from its approximately linear IA-like scaling and gradually approaches a nearly constant frequency. Because $\Omega_i$ is smallest near the weak-field center and increases away from it, this change occurs at smaller $k$ near $X=0$ and at larger $k$ in the outer parts of the sheet.

The position within the current sheet therefore affects both the frequency of the low-$k$ IA-like branch and the wavenumber at which ion gyromotion begins to modify it. Near the sheet center, the branch has a lower frequency and leaves its acoustic behavior at a smaller $k$. Farther from the center, the frequency is higher and the low-$k$ acoustic behavior extends to larger wavenumbers. The nearly constant high-$k$ parts of these curves should not be interpreted as IA-like waves; they represent the continuation of the lower electrostatic root as its character becomes increasingly controlled by ion gyromotion.

\subsubsection{Effect of Pair Multiplicity}
\label{subsec:kappa-dependence}

Figure~\ref{fig:frequency-vs-k}(b) shows the variation of $\omega_{\rm IA}$ with wavenumber $k$ at the current-sheet center, $X=0$, for pair multiplicities from $\kappa=1\times10^5$ to $6\times10^5$. At small $k$, all curves increase approximately linearly, as expected for the IA-like part of the lower branch. Increasing $\kappa$ shifts the curves to lower frequencies, with the largest separation occurring before the branch approaches its high-wavenumber limit.

This decrease follows from the way the pair population enters both the equilibrium and the electrostatic response. A larger $\kappa$ increases the electron and positron densities and strengthens pair shielding. At the same time, pressure balance distributes the available thermal pressure over a larger pair population, which lowers the pair temperature and, through $T_{i0}=\tau_iT_{\ell0}$, also lowers the ion temperature. Both effects reduce the effective IA speed $C_{\rm IA}$ and therefore lower $\omega_{\rm IA}$ over the acoustic part of the branch.

At larger $k$, the curves approach nearly the same limiting frequency. In this range, the lower branch is increasingly controlled by ion gyromotion rather than by the pair-dependent acoustic response. Since the local ion gyrofrequency and propagation angle are unchanged in this comparison, the dependence on $\kappa$ becomes much weaker near the high-wavenumber limit.

Figure~\ref{fig:frequency-vs-k}(b) therefore shows that pair multiplicity mainly controls the low and intermediate-wavenumber IA-like waves. A denser pair background lowers the wave frequency through stronger shielding and the associated change in the pressure-balanced temperature, while having much less effect once the lower branch approaches its ion-gyromotion dominated limit.

\subsection{Effect of propagation angle}
\label{subsec:angular-dependence}

Figure~\ref{fig:frequency-vs-k}(c) shows the effect of angle of propagation $\theta$ with respect to the local magnetic field. The angle is varied from parallel propagation, $\theta=0^\circ$, to strongly oblique propagation, $\theta=65^\circ$. At a fixed wavenumber, the wave frequency decreases as $\theta$ is increased.
For parallel propagation (i.e., $k_\perp=0$), the compressive ion motion is directed entirely along the magnetic field. Then the perpendicular Lorentz force does not affect the IA-like wave dynamics, and the dispersion relation separates into the purely acoustic and ion-cyclotron solutions, $\omega_{\rm ac}=k\,C_{\rm IA}$ and $\omega_{\rm ci}=\Omega_i$, respectively.

The two parallel solutions cross at the wavenumber $k_{\rm tr}$ defined by $k_{\rm tr}C_{\rm IA}(k_{\rm tr})=\Omega_i$. Below this point, the green lower-root curve for $\omega_{\rm IA }\to \omega_{\rm ac}$, while the red upper-root curve remains at the ion-cyclotron frequency, i.e., $\omega_{\rm IC }=\omega_{\rm ci}$. Above the crossing, their frequency ordering is reversed: the lower root follows the constant ion-cyclotron solution, i.e., $\omega_{\rm IA }\to\omega_{\rm ci}$, whereas $\omega_{\rm IC }\to\omega_{\rm ac}$. The horizontal part of the green curve should therefore not be interpreted as a nondispersive IA wave. It represents the ion-cyclotron solution after the two parallel branches cross.

For oblique propagation, the acoustic and ion-cyclotron behavior are coupled through the perpendicular ion motion, so the exact crossing is replaced by a smooth bending of the curves over approximately the same range of wavenumbers. This bending marks the gradual change from a mainly acoustic behavior at small $k$ to a response increasingly limited by ion gyromotion at larger $k$.

As $\theta$ approaches $90^\circ$, the parallel wavenumber $k_\parallel=k\cos\theta$ tends to zero, and the frequency of the lower branch correspondingly vanishes. Within the present model, the IA-like branch therefore requires a finite component of the wave vector along the magnetic field. This angular dependence also shows why the perpendicular velocity components $u_x$ and $u_y$ must be retained in the fluid derivation: omitting them would remove the magnetic modification of the oblique branch.

\subsection{$P$--$\dot P$ Dependence and the Acoustic--Cyclotron
Transition}

Figure~\ref{fig:P-Pdot-map} extends the local current-sheet calculation over the pulsar $P$--$\dot P$ plane. At each point, the surface and light-cylinder magnetic fields and the Goldreich--Julian density are recalculated from the relations in Sec.~\ref{sec:model}. All dimensionless current-sheet parameters are otherwise held fixed, with $r=50R_{\rm LC}$ and $X/\Delta=0.1$. The calculation is therefore a controlled comparison of the effect of the host-pulsar spin parameters rather than a source-by-source model of the current sheet.

Panel~(a) shows the lower electrostatic root $\omega_{\rm IA}$ at the fixed physical wavenumber $k_{\rm ref}=5.0\times10^{-5}\ {\rm rad\,cm^{-1}}$, corresponding to $\lambda_{\rm ref}=2\pi/k_{\rm ref}\simeq1.26\times10^5$ cm, for parallel propagation. This illustrative wavenumber is chosen so that the acoustic--cyclotron transition crosses the range occupied by several young energetic pulsars. For $\theta=0^\circ$, the dispersion relation factorizes exactly into $\omega_{\rm ac}=kC_{\rm IA}$ and $\omega_{\rm ci}=\Omega_i$. The transition wavenumber is therefore defined by
\begin{equation}
k_{\rm tr}C_{\rm IA}(k_{\rm tr})=\Omega_i .
\label{eq:ktr_ppdot}
\end{equation}
The dashed contour in Fig.~\ref{fig:P-Pdot-map} marks $k_{\rm tr}=k_{\rm ref}$. For $k_{\rm ref}<k_{\rm tr}$, the lower root is the acoustic solution and can be identified as IA-like, subject to the additional existence conditions of Sec.~\ref{sec:existence}.  For $k_{\rm ref}>k_{\rm tr}$, the two parallel branches have already exchanged their frequency ordering and the lower root is the ion-cyclotron solution. The dotted region in panel~(a) marks this latter domain.

The strong period dependence seen in panel~(a) follows directly from the pulsar-linked equilibrium. At fixed $r/R_{\rm LC}$, $\Gamma_w$,
$X/\Delta$, and dimensionless sheet parameters,
\[
B_0\propto \dot P^{1/2}P^{-5/2},
\qquad
n_{\rm GJ}\propto\dot P^{1/2}P^{-7/2}.
\]
Pressure balance then gives
$T_i\propto B_{\rm up}^2/n_{\rm GJ}
\propto\dot P^{1/2}P^{-3/2}$. In the pair-rich regime considered here, it is observed that the long-wavelength acoustic speed is dominated by warm-ion pressure, so $C_{\rm IA}\simeq c_{s,i}\propto  \dot P^{1/4}P^{-3/4}$.
Consequently, on the acoustic side of the transition,
\begin{equation}
\omega_{\rm IA}(k_{\rm ref})
\simeq k_{\rm ref}C_{\rm IA}
\propto\dot P^{1/4}P^{-3/4},\nonumber
\end{equation}
whereas after the branch exchange the lower root follows
\begin{equation}
\omega_{\rm IA}(k_{\rm ref})
=\Omega_i
\propto\dot P^{1/2}P^{-5/2}.\nonumber
\end{equation}
The lower-root frequency therefore increases toward shorter periods in both regions, but considerably more steeply once the selected wavenumber lies on the ion-cyclotron branch.

Panel~(b) maps $k_{\rm tr}$ itself. In the same warm-ion-pressure-dominated limit, 
\begin{equation}
k_{\rm tr}\simeq\frac{\Omega_i}{C_{\rm IA}}
\propto\dot P^{1/4}P^{-7/4},\nonumber
\end{equation}
which explains the much stronger dependence on $P$ than on $\dot P$ and the rapid decrease of the transition wavenumber toward longer-period pulsars. The contour $k_{\rm tr}=k_{\rm ref}$ therefore separates pulsars for which the chosen physical scale samples the acoustic part of the lower branch from those for which it samples the ion-cyclotron part.

For the common current-sheet parameters adopted here, J0058--7218 and J0537--6910 lie on the acoustic side of this transition, the Crab lies close to it, and several longer-period sources lie beyond it. These locations should not be interpreted as source-specific predictions of IA-wave activity: changing $\Gamma_w$, composition, guide field, sheet thickness, temperature partition, or position within the sheet also changes the transition scale.
A lower root that is acoustic at $k_{\rm ref}$ is physically admissible within the present model only if
\begin{equation}
k_{\rm low}(r,X)<k_{\rm ref}<k_{\rm high}(r,X),\nonumber
\end{equation}
together with the position-dependent validity and current-closure conditions of Sec.~\ref{subsec:combined_existence}. The $P$--$\dot P$ map therefore determines which formal branch is sampled at a fixed physical wavelength, whereas the complete existence analysis determines whether that
branch can be treated as a local IA-like fluid mode.

\subsection{Restoring force}
\label{subsec:restoring-force}

In this model, ions provide the necessary inertia, and the pairs dominate the electrostatic shielding, but it does not mean that the $\omega_{\rm IA}$ branch is pair-shielded in the usual sense. In the long-wavelength limit $k\lambda_{\rm pair}\ll1$, Eq.~(\ref{eq:Cs_effective_def}) separates into an ion-pressure term and a
pair-shielded term whose ratio is fixed by the local charge ratio alone,
\begin{equation}
\frac{c_{s,i}^{2}}{\omega_{pi}^{2}\lambda_{\rm pair}^{2}}
=\frac{\gamma_i\tau_i\,(1+2\alpha)}{Z_i}.
\label{eq:ratio}
\end{equation}
With $\gamma_i=5/3$, $\tau_i=0.1$, $Z_i=1$, and $\alpha\simeq1.1\times10^{3}$ at the sheet center, this ratio is $\approx3.7\times10^{2}$: the pair-shielded contribution supplies less than $0.3\%$ of $C_{\rm IA}^{2}$, and rather less still in the outer sheet where $\alpha\to10^{4}$. The low-frequency compressive mode $\omega_{\rm IA}$ is therefore supported almost entirely by warm ion pressure, with Debye shielding entering as a small correction.

Equation~(\ref{eq:ratio}) also gives the condition under which this changes. The pair-shielded term dominates only for
\begin{equation}
\alpha<\frac{Z_i-\gamma_i\tau_i}{2\gamma_i\tau_i}\simeq2.5 ,
\end{equation}
nearly three orders of magnitude below the sheet center value obtained with $A_i/A_p=10$. Achieving this would require either $\kappa/\eta\lesssim25$, far below the multiplicities expected from pair cascades \citep{TimokhinHarding2015,TimokhinHarding2019}, or sheet center compression ratio $\mathcal{C}_i(0)/\mathcal{C}_p(0)\gtrsim10^{3}$.

This strengthens the core argument rather than weakening it. Ion loading is necessary, because a mass-symmetric pair plasma has no heavy inertial species \citep{Tsytovich1970,Verheest2000}. It is not sufficient, because at realistic pair multiplicities the electrostatic restoring force is screened out and what survives is ion acoustic mode carried by the pressure-balanced $T_{i0}$. 
The branch remains a diagnostic of ion loading, but through $c_{s,i}$ rather than through $\lambda_{\rm pair}$, and Eq.~(\ref{eq:ratio}) provides the test that separates the two.

\section{Discussion}
\label{sec:discussion}
    
\subsection{Physical interpretation of the IA-like window}
\label{subsec:discussion_window}
    
The analysis in this paper reveals that, in pulsar wind current sheet, IA-like modes exist only under specific existence conditions. The pressure-balanced current-sheet equilibrium fixes the densities and temperatures and therefore also sets $C_{\rm IA}$, $\lambda_{\rm pair}$, $\omega_{pi}$, $\Omega_i$, and $\rho_i$. The IA-like branch is consequently restricted to the finite interval $k_{\rm low}(r,X)<k<k_{\rm high}(r,X)$, together with the position-dependent ion temperature, speed, and current-closure conditions derived in Sec.~\ref{subsec:combined_existence}. The dispersion relation alone is therefore not sufficient to determine where the mode can be described consistently by the present fluid model.

At small wavenumber, the local treatment fails when the wavelength becomes too large compared with the spatial variation of the equilibrium or when the wave period becomes comparable with the adopted sheet-evolution or advection time. At large wavenumber, the lower root departs from its acoustic character and the assumptions of  strong ion magnetization, long perpendicular wavelength, and pair shielding become increasingly restrictive. For the fiducial parameters, the lower-$k$ range is limited mainly by the spatial variation of the sheet, while at high-$k$ the spatial pair-shielding condition becomes restrictive before the temporal pair-response condition.

The sheet center shows why stronger ion enrichment does not necessarily produce a wider IA-like range. In the adopted density profile, the ion density is largest near $X=0$, so the ions contribute more strongly to the local wave dynamics. At the same time, the reversing magnetic field falls to zero there and only the guide field remains. This lowers $\Omega_i$ and increases $\rho_i$, so the allowed range set by ion magnetization and the ion gyroradius becomes narrower near the center. Thus, the IA-like mode existence domain therefore depends on a balance between stronger ion loading and the weaker magnetic field across the sheet.

\subsection{Composition and the restoring force}
\label{subsec:discussion_composition}

For the existence of IA-like mode, presence of ions is essential because they provide the heavy inertial component that is absent in a symmetric electron--positron plasma in the pulsar wind. Their presence alone, however, does not determine the character of the restoring force for IA-like waves. The restoring force decomposition derived in Sec.~\ref{subsec:restoring-force} shows that, for the pair-dominated sheet, the warm-ion pressure term is much larger than the pair-shielded contribution to $C_{\rm IA}^2$. 
The resulting mode is therefore an ion-loaded compressive mode whose phase speed is set mainly by the pressure-balanced ion temperature, while the pair population still controls the electrostatic shielding scale.
Furthermore, increasing $\kappa$ increases the pair density and redistributes the available thermal pressure among more particles. 
The pair temperature decreases, and $T_{i0}=\tau_iT_{\ell0}$ decreases with it.
The reduction of $T_{i0}$ lowers the warm-ion contribution to $C_{\rm IA}$, while the larger pair density changes $\lambda_{\rm pair}$. 
The decrease of $\omega_{\rm IA}$ with increasing $\kappa$ therefore follows from the coupled density and temperature changes imposed by the equilibrium rather than from a single shielding effect.

\subsection{Dependence on the host pulsar}
\label{subsec:discussion_pulsar}

The local wave properties are linked to the host pulsar through $P$ and $\dot P$, which determine the magnetic-field and Goldreich--Julian density scales used to construct the comoving current-sheet equilibrium. The $P$--$\dot P$ maps show that these parameters affect both the frequency evaluated at a fixed physical wavenumber and the transition scale $k_{\rm tr}$. 
Two pulsars evaluated at the same $k_{\rm ref}$ can place that wavenumber on different sides of the acoustic--cyclotron transition.

For the fixed current-sheet parameters adopted in Figure~\ref{fig:P-Pdot-map}, the variation is controlled mainly by the spin period, with a weaker dependence on $\dot P$ over the plotted range. J0058--7218 and J0537--6910 lie on the acoustic side of the transition at the adopted $k_{\rm ref}$, the Crab lies close to the transition, and several longer-period sources lie beyond it. This comparison does not make branch identity a function of $P$ and $\dot P$ alone: $k_{\rm tr}$ also depends on the assumed wind Lorentz factor, composition, temperature partition, guide field, compression, and position within the sheet. The $P$--$\dot P$ calculation instead shows how the host pulsar changes the local wave scales when those current-sheet properties are held fixed.

The present electrostatic calculation also remains separated from ion-kinetic electromagnetic physics by its adopted ordering. In particular, the ion-fluid treatment requires $k_\perp\rho_i\leq0.1$. The analysis therefore does not extend into the ion-gyroradius regime where finite-Larmor-radius and kinetic Alfv\'en effects become important. A kinetic treatment would be required to follow any connection between these regimes.

\subsection{Limits of the fluid description}
\label{subsec:discussion_kinetic}

The present theory determines the real frequency and the range of parameters for which the IA-like identification is consistent with the adopted local fluid equations. It does not determine the imaginary part of the frequency. Collisionless damping or growth depends on the particle distribution functions, relative drifts, and resonant particles, none of which are specified by the present pressure closure \citep{Gary1993,Stix1992}.
This limitation is particularly relevant because the acoustic phase speed can be comparable with characteristic ion thermal speeds in the warm-ion regime. Ion resonances may therefore be important, but their strength and even the net sign of the imaginary frequency require a kinetic calculation. The same applies to possible free energy from relative pair--ion drifts or reconnection flows.

A useful next step is therefore to retain the pressure-balanced current-sheet equilibrium developed here while replacing the static pair susceptibility and warm-ion fluid closure by kinetic susceptibilities. The resulting complex frequency could then be compared directly with the advection and sheet-evolution times already defined in this work. This would separate three questions that are distinct in a collisionless current sheet: whether the branch exists, whether it lies within the local current-sheet ordering, and whether kinetic damping or growth allows it to persist over the available time.

\section{Conclusions}
\label{sec:conclusions}

We have developed a theory for IA-like modes in an ion-loaded, pressure-balanced pulsar-wind current sheet. The local plasma properties are linked to the pulsar through $P$, $\dot P$, and the wind Lorentz factor, while the magnetic-field and density structure across the sheet determine the thermal state through pressure balance. The pair and ion temperatures are therefore fixed by the current-sheet equilibrium rather than introduced as independent parameters in the dispersion relation.

The resulting warm-ion electrostatic dispersion relation contains coupled acoustic and IC branches. For parallel propagation, the two solutions cross at $k_{\rm tr}C_{\rm IA}(k_{\rm tr})=\Omega_i$. At finite obliquity, the acoustic and IC behaviors are coupled, and the exact crossing is replaced by a gradual transition. The lower root can be identified as IA-like only over the part of the spectrum where it remains mainly acoustic and stays within the conditions of the local fluid model.

A main result is the finite wavenumber interval $k_{\rm low}(r,X)<k<k_{\rm high}(r,X)$. The lower boundary is set by the local-inhomogeneity, quasi-static, and advection-time conditions, whereas the upper boundary follows from branch identity, ion magnetization, pair response, and the finite ion gyroradius. The sheet center contains the strongest relative ion enrichment in the adopted model, but it also has the weakest magnetic field. The corresponding decrease of $\Omega_i$ and increase of $\rho_i$ narrow the allowed high-$k$ range near the center. Stronger ion enrichment therefore does not necessarily produce a wider IA-like domain.

The parameter studies show how the branch changes across the current sheet and with the plasma conditions. Moving away from the sheet center increases the frequency and shifts the departure from the IA-like regime to larger $k$, mainly because the local magnetic field and $\Omega_i$ increase. Increasing the pair multiplicity $\kappa$ lowers the frequency over the low- and intermediate-$k$ part of the branch through the combined changes in density, shielding, and the pressure-balanced temperature. The dependence on $\kappa$ becomes weaker at larger $k$, where ion gyromotion has a larger influence on the lower root. Increasing the propagation angle $\theta$ also lowers the IA-like frequency because $k_\parallel=k\cos\theta$ decreases. For parallel propagation the acoustic and IC solutions cross exactly, whereas at finite obliquity their behaviors become smoothly coupled.

The restoring-force balance gives a further physical interpretation of these trends. For the fiducial pair-dominated parameters, the low-frequency branch is supported mainly by warm-ion pressure. The pairs remain important because they set the electrostatic shielding scale and also enter the pressure balance that fixes the local temperature, but their direct shielded contribution to $C_{\rm IA}^2$ is small for the large pair-to-ion charge ratios considered here. The IA-like dynamics therefore depend on both the presence of ions and the thermal state produced by the pressure-balanced sheet.

The $P$--$\dot P$ calculation shows that the host pulsar changes both the wave frequency at a fixed physical scale and the acoustic--IC transition wavenumber. At the same physical wavelength, the lower root can remain on the IA-like branch for one pulsar but lie beyond the acoustic--IC transition for another, even when the other current-sheet parameters are kept fixed. The pulsar spin parameters therefore influence not only the characteristic wave frequency but also the range of wavenumbers over which the lower branch remains IA-like.

Overall, the model identifies the locations and wavenumbers where an ion-loaded electrostatic compressive branch can remain IA-like inside a pulsar-wind current sheet. It also shows how this range changes with position in the sheet, pair multiplicity, propagation angle, magnetic geometry, and the host pulsar. A kinetic treatment of the same pressure-balanced background can extend this picture by determining the damping or growth of the branch and how long it can persist in the current sheet.

\begin{acknowledgments}
M.S. gratefully acknowledges financial support from the Basic Scientific Research Fund for Central Universities, China (Grant No. 2682025CX094). SL acknowledge the support by National Natural Science Foundation of China under the Grant No. 12375103.
\end{acknowledgments}

\bibliography{References}
\bibliographystyle{aasjournalv7}

\end{document}